%% file: DESY-10-047.tex
\documentclass[zpreprint,zbstdefault]{zeus_paper}
\usepackage[english]{babel}

\input  zeus_def.tex
\input  def.tex
\input  DESY-10-047-cit.tex
\begin{document}
\include{DESY-10-047-tit}
\include{DESY-10-047-aut}
\include{DESY-10-047-txt}
\include{DESY-10-047-ref}
\include{DESY-10-047-tab}

\include{DESY-10-047-fig}

%
%
\end{document}

%% file: zeus_def.tex
\newcommand{\ZpsrapB}{%
The pseudorapidity is defined as $\eta=-\ln\left(\tan\frac{\theta}{2}\right)$,
where the polar angle, $\theta$, is measured with respect to the $Z$ axis.
The ZEUS coordinate system is a right-handed Cartesian system, with the $Z$
axis pointing in the proton beam direction, referred to as the ``forward
direction'', and the $X$ axis pointing towards the centre of HERA.
The coordinate origin is at the nominal interaction point.\xspace}

\newcommand{\Zdetdesc}{%
A detailed description of the ZEUS detector can be found 
elsewhere~\cite{zeus:1993:bluebook}. A brief outline of the 
components that are most relevant for this analysis is given
below.\xspace}

\newcommand{\ZctddescB}{%
Charged particles were tracked in the central tracking detector (CTD)~\citeCTD,
which operated in a magnetic field of $1.43\Tesla$ provided by a thin 
superconducting coil. The CTD consisted of 72~cylindrical drift chamber 
layers, organised in 9~superlayers covering the polar-angle region 
\mbox{$15^\circ<\theta<164^\circ$}. The transverse-momentum resolution for
full-length tracks is $\sigma(p_T)/p_T=0.0058p_T\oplus0.0065\oplus0.0014/p_T$,
with $p_T$ in $\Gev$.}

\newcommand{\ZcaldescB}{%
The high-resolution uranium--scintillator calorimeter (CAL)~\citeCAL consisted 
of three parts: the forward (FCAL), the barrel (BCAL) and the rear (RCAL)
calorimeters. Each part was subdivided transversely into towers and
longitudinally into one electromagnetic section and either one (in RCAL)
or two (in BCAL and FCAL) hadronic sections. 
The CAL energy resolutions, as measured under
test-beam conditions, are $\sigma(E)/E=0.18/\sqrt{E}$ for electrons and
$\sigma(E)/E=0.35/\sqrt{E}$ for hadrons, with $E$ in $\Gev$.}

\chardef\usc=95
\chardef\til=126
\catcode`\@=11 
\DeclareRobustCommand\xdotspace{\futurelet\@let@token\@xdotspace}
\def\@xdotspace{%
  \ifx\@let@token.\else
  \ifx\@let@token\bgroup.\else
  \ifx\@let@token\egroup.\else
  \ifx\@let@token\/.\else
  \ifx\@let@token\ .\else
  \ifx\@let@token~.\else
  \ifx\@let@token!.\else
  \ifx\@let@token,.\else
  \ifx\@let@token:.\else
  \ifx\@let@token;.\else
  \ifx\@let@token?.\else
  \ifx\@let@token/.\else
  \ifx\@let@token'.\else
  \ifx\@let@token).\else
  \ifx\@let@token-.\else
  \ifx\@let@token\@xobeysp.\else
  \ifx\@let@token\space.\else
  \ifx\@let@token\@sptoken.\else
   .\space
   \fi\fi\fi\fi\fi\fi\fi\fi\fi\fi\fi\fi\fi\fi\fi\fi\fi\fi}
\catcode`\@=12 

\newcommand{\stru}[2]{%
   \relax\ifmmode\hbox{\vrule height#1 depth#2 width0pt}%
   \else\vrule height#1 depth#2 width0pt\fi}

\newcommand{\Ronum}[1]{\uppercase\expandafter{\romannumeral#1}}
\newcommand{\ronum}[1]{\expandafter{\romannumeral#1}}
\DeclareRobustCommand{\LaTeXZ}{%
  \LaTeX\kern-.05em4\kern-.1em
  {\raisebox{-0.2ex}{$\scriptstyle\text{ZEUS}$}}\xspace}

\DeclareMathAlphabet{\mathbf}{OT1}{cmr}{bx}{sl}
\newcommand{\eVdist}{\kern-0.06667em}

\newcommand{\Gev}{{\text{Ge}\eVdist\text{V\/}}}

\newcommand{\gev}{{\,\text{Ge}\eVdist\text{V\/}}}

\newcommand{\pb}{\,\text{pb}}

\newcommand{\pbi}{\,\text{pb}^{-1}}

\newcommand{\cm}{\,\text{cm}}

\newcommand{\Tesla}{\,\text{T}}

\newcommand{\slashfrac}[2]{%
  \raisebox{0.5ex}{\ensuremath #1}\kern-0.12em/\kern-0.08em
  \raisebox{-.8ex}{\ensuremath #2}}

\newcommand{\sqr}[3]{%
    {\vcenter{\hrule height.#3ex\hbox{\vrule width.#2ex height#1ex
     \kern#1ex\vrule width.#3ex}\hrule height.#2ex}}}

\newcommand{\widebar}[1]{%
   \mkern1.5mu\overline{\mkern-1.5mu#1\mkern-1.mu}\mkern1.mu}
\catcode`\@=11 
\newcommand{\parenbar}{\mathpalette\p@renb@r}
\def\p@renb@r#1#2{\vbox{%
  \ifx#1\scriptscriptstyle \dimen@.7em\dimen@ii.2em\else
  \ifx#1\scriptstyle \dimen@.8em\dimen@ii.25em\else
  \dimen@1em\dimen@ii.4em\fi\fi \offinterlineskip
  \ialign{\hfill##\hfill\cr
    \vbox{\hrule width\dimen@ii}\cr
    \noalign{\vskip-.3ex}%
    \hbox to\dimen@{$\mathchar300\hfil\mathchar301$}\cr
    \noalign{\vskip-.3ex}%
    $#1#2$\cr}}}
\catcode`\@=12 

\newcommand{\pbar}{\widebar{p}}

\newcommand{\bbar}{\widebar{b}}

\newcommand{\IP}{{\rm I$\kern-0.01667em$P}\xspace}

\mathchardef\qsm=63
\mathchardef\pls=43
\mathchardef\mns=512
\mathchardef\plm=518
\mathchardef\eql=61
\mathchardef\smallleft=300
\mathchardef\smallright=301
\mathchardef\les=316
\mathchardef\gre=318
\mathchardef\leq=532
\mathchardef\grq=533

\catcode`\@=11 
\newcounter{pict@width}
\newcounter{pict@height}
\newlength{\pict@scale}
\newcommand{\psfigadd}[4]{%
\setcounter{pict@width}{1*\ratio{#2+\pict@scale/2}{\pict@scale}}
\setcounter{pict@height}{1*\ratio{#3+\pict@scale/2}{\pict@scale}}
\setlength{\unitlength}{\pict@scale}
\hbox to #2{\hspace{-\fill}\begin{picture}(\thepict@width,\thepict@height)
\put(0,0){\psfig{figure=#1,width=#2,height=#3,clip=}}
\SetScale{0.283466457}
\SetWidth{1.763889}
{#4}
\end{picture}}
}
\newcounter{pict@widthfst}
\newcounter{pict@widthscd}
\newcounter{pict@widthtot}
\newcommand{\psfigaddtwo}[7]{%
\setcounter{pict@widthfst}{1*\ratio{#2+\pict@scale/2}{\pict@scale}}
\setcounter{pict@widthscd}{1*\ratio{#2+#4+\pict@scale/2}{\pict@scale}}
\setcounter{pict@widthtot}{1*\ratio{#2+#4+#6+\pict@scale/2}{\pict@scale}}
\setcounter{pict@height}{1*\ratio{#3+\pict@scale/2}{\pict@scale}}
\setlength{\unitlength}{\pict@scale}
\hbox{\hspace{-\fill}\begin{picture}(\thepict@widthtot,\thepict@height)
\put(0,0){\psfig{figure=#1,width=#2,height=#3,clip=}}
\put(\thepict@widthscd,0){\psfig{figure=#5,width=#6,height=#3,clip=}}
\SetScale{0.283466457}
\SetWidth{1.763889}
{#7}
\end{picture}}
}
\newcommand{\psfigror}[4]{%
\setcounter{pict@width}{1*\ratio{#2+\pict@scale/2}{\pict@scale}}
\setcounter{pict@height}{1*\ratio{#3+\pict@scale/2}{\pict@scale}}
\setlength{\unitlength}{\pict@scale}
\hbox{\begin{picture}(\thepict@width,\thepict@height)
\put(0,\thepict@height){\psfig{figure=#1,width=#3,height=#2,clip=,angle=270}}
\SetScale{0.283466457}
\SetWidth{1.763889}
{#4}
\end{picture}}
}
\newcommand{\psfigrol}[4]{%
\setcounter{pict@width}{1*\ratio{#2+\pict@scale/2}{\pict@scale}}
\setcounter{pict@height}{1*\ratio{#3+\pict@scale/2}{\pict@scale}}
\setlength{\unitlength}{\pict@scale}
\hbox{\begin{picture}(\thepict@width,\thepict@height)
\put(0,0){\psfig{figure=#1,width=#3,height=#2,clip=,angle=90}}
\SetScale{0.283466457}
\SetWidth{1.763889}
{#4}
\end{picture}}
}
\catcode`\@=12 
\newlength\listtextwidth

\catcode`\@=11 
\newlength{\@tabfninsert}
\newlength{\@tabfnwidth}
\newcommand{\tabfootnote}[2]{%
  \setlength{\@tabfninsert}{0.8em}
  \setlength{\@tabfnwidth}{\textwidth}
  \addtolength{\@tabfnwidth}{-\@tabfninsert}
  \addtolength{\@tabfnwidth}{-0.4em}
  \noindent\makebox[\@tabfninsert][r]{\footnotesize$^{#1}$\hfil}\hfill%
  \parbox[t]{\@tabfnwidth}{\footnotesize #2\hfill}}
\catcode`\@=12 

%% file: def.tex
\newcommand{\geant}{\textsc{Geant}\xspace}

\newcommand{\rapgap}{\textsc{Rapgap}\xspace}
\newcommand{\ariadne}{\textsc{Ariadne}\xspace}

\newcommand{\hvqdis}{\textsc{Hvqdis}\xspace}

\newcommand{\ktclus}{\textsc{Ktclus}\xspace}

\newcommand{\bbbar}{\ensuremath{b\bar{b}}\xspace}
\newcommand{\ccbar}{\ensuremath{c\bar{c}}\xspace}

\newcommand{\alphas}{\ensuremath{\alpha_{\mathrm{s}}}\xspace}

\newcommand{\ptrel}{\ensuremath{p_T^{\mathrm{rel}}}\xspace}

\newcommand{\calet}{\ensuremath{E_T^{\mathrm{cal}}}\xspace}
\newcommand{\caletcorr}{\ensuremath{E_T}\xspace}

\newcommand{\jpsi}{\ensuremath{J/\psi}\xspace}

\newcommand{\Ftwob}{\ensuremath{F_2^{b\bbar}}\xspace}

%% file: DESY-10-047-cit.tex
\def\citeCTD{{\cite{%
nim:a279:290,*npps:b32:181,*nim:a338:254%
}}\xspace}
\def\citeCAL{{\cite{%
nim:a309:77,*nim:a309:101,*nim:a321:356,*nim:a336:23%
}}\xspace}

%% file: DESY-10-047-tit.tex
\prepnum{DESY--10--047}

\title{
Measurement of beauty production in DIS and {\boldmath \Ftwob} extraction at ZEUS
}                                                       
                    
\author{ZEUS Collaboration}
\date{April 30, 2010}

\abstract{
Beauty production in deep inelastic scattering with events in which a muon 
and a jet are observed in the final state has been measured with the ZEUS 
detector at HERA using an integrated luminosity of $114\pbi$. The fraction of 
events with beauty quarks in the data was determined using the distribution 
of the transverse momentum of the muon relative to the jet. The cross section 
for beauty production was measured in the kinematic range of photon virtuality,
$Q^2 > 2 \gev^2$, and inelasticity, $0.05\,<\,y\,<\,0.7$, with the requirement 
of a muon and a jet.
Total and differential cross sections are presented and compared to QCD 
predictions. The beauty contribution to the structure function $F_2$ was 
extracted and is compared to theoretical predictions. 
}

\makezeustitle

%% file: DESY-10-047-aut.tex
%
%
%
%

\def\3{\ss}
\pagenumbering{Roman}
                                                   %
\begin{center}
{                      \Large  The ZEUS Collaboration              }
\end{center}

{\small


{\mbox H.~Abramowicz$^{44, ad}$, }
{\mbox I.~Abt$^{34}$, }
{\mbox L.~Adamczyk$^{13}$, }
{\mbox M.~Adamus$^{53}$, }
{\mbox R.~Aggarwal$^{7}$, }
{\mbox S.~Antonelli$^{4}$, }
{\mbox P.~Antonioli$^{3}$, }
{\mbox A.~Antonov$^{32}$, }
{\mbox M.~Arneodo$^{49}$, }
{\mbox V.~Aushev$^{26, y}$
% ...
}
\newpage


\makebox[3em]{$^{1}$}
\begin{minipage}[t]{14cm}
{\it Argonne National Laboratory, Argonne, Illinois 60439-4815, USA}~$^{A}$

\end{minipage}\\
\makebox[3em]{$^{2}$}
\begin{minipage}[t]{14cm}
{\it Andrews University, Berrien Springs, Michigan 49104-0380, USA}

\end{minipage}\\
\makebox[3em]{$^{3}$}
\begin{minipage}[t]{14cm}
{\it INFN Bologna, Bologna, Italy}~$^{B}$

\end{minipage}\\
\makebox[3em]{$^{4}$}
\begin{minipage}[t]{14cm}
{\it University and INFN Bologna, Bologna, Italy}~$^{B}$

\end{minipage}\\
\makebox[3em]{$^{5}$}
\begin{minipage}[t]{14cm}
{\it Physikalisches Institut der Universit\"at Bonn,
Bonn, Germany}~$^{C}$

\end{minipage}\\
\makebox[3em]{$^{6}$}
\begin{minipage}[t]{14cm}
{\it H.H.~Wills Physics Laboratory, University of Bristol,
Bristol, United Kingdom}~$^{D}$

\end{minipage}\\
\makebox[3em]{$^{7}$}
\begin{minipage}[t]{14cm}
{\it Panjab University, Department of Physics, Chandigarh, India}

\end{minipage}\\
\makebox[3em]{$^{8}$}
\begin{minipage}[t]{14cm}
{\it Calabria University,
Physics Department and INFN, Cosenza, Italy}~$^{B}$

\end{minipage}\\
\makebox[3em]{$^{9}$}
\begin{minipage}[t]{14cm}
{\it Institute for Universe and Elementary Particles, Chonnam National University,\\
Kwangju, South Korea}

\end{minipage}\\
\makebox[3em]{$^{10}$}
\begin{minipage}[t]{14cm}
{\it Jabatan Fizik, Universiti Malaya, 50603 Kuala Lumpur, Malaysia}~$^{E}$

\end{minipage}\\
\makebox[3em]{$^{11}$}
\begin{minipage}[t]{14cm}
{\it Nevis Laboratories, Columbia University, Irvington on Hudson,
New York 10027, USA}~$^{F}$

\end{minipage}\\
\makebox[3em]{$^{12}$}
\begin{minipage}[t]{14cm}
{\it The Henryk Niewodniczanski Institute of Nuclear Physics, Polish Academy of Sciences,\\
Cracow, Poland}~$^{G}$

\end{minipage}\\
\makebox[3em]{$^{13}$}
\begin{minipage}[t]{14cm}
{\it Faculty of Physics and Applied Computer Science, AGH-University of Science and \\
Technology, Cracow, Poland}~$^{H}$

\end{minipage}\\
\makebox[3em]{$^{14}$}
\begin{minipage}[t]{14cm}
{\it Department of Physics, Jagellonian University, Cracow, Poland}

\end{minipage}\\
\makebox[3em]{$^{15}$}
\begin{minipage}[t]{14cm}
{\it Deutsches Elektronen-Synchrotron DESY, Hamburg, Germany}

\end{minipage}\\
\makebox[3em]{$^{16}$}
\begin{minipage}[t]{14cm}
{\it Deutsches Elektronen-Synchrotron DESY, Zeuthen, Germany}

\end{minipage}\\
\makebox[3em]{$^{17}$}
\begin{minipage}[t]{14cm}
{\it INFN Florence, Florence, Italy}~$^{B}$

\end{minipage}\\
\makebox[3em]{$^{18}$}
\begin{minipage}[t]{14cm}
{\it University and INFN Florence, Florence, Italy}~$^{B}$

\end{minipage}\\
\makebox[3em]{$^{19}$}
\begin{minipage}[t]{14cm}
{\it Fakult\"at f\"ur Physik der Universit\"at Freiburg i.Br.,
Freiburg i.Br., Germany}~$^{C}$

\end{minipage}\\
\makebox[3em]{$^{20}$}
\begin{minipage}[t]{14cm}
{\it Department of Physics and Astronomy, University of Glasgow,
Glasgow, United Kingdom}~$^{D}$

\end{minipage}\\
\makebox[3em]{$^{21}$}
\begin{minipage}[t]{14cm}
{\it Department of Engineering in Management and Finance, Univ. of
the Aegean, Chios, Greece}

\end{minipage}\\
\makebox[3em]{$^{22}$}
\begin{minipage}[t]{14cm}
{\it Hamburg University, Institute of Exp. Physics, Hamburg,
Germany}~$^{C}$

\end{minipage}\\
\makebox[3em]{$^{23}$}
\begin{minipage}[t]{14cm}
{\it Imperial College London, High Energy Nuclear Physics Group,
London, United Kingdom}~$^{D}$

\end{minipage}\\
\makebox[3em]{$^{24}$}
\begin{minipage}[t]{14cm}
{\it Institute of Particle and Nuclear Studies, KEK,
Tsukuba, Japan}~$^{I}$

\end{minipage}\\
\makebox[3em]{$^{25}$}
\begin{minipage}[t]{14cm}
{\it Institute of Physics and Technology of Ministry of Education and
Science of Kazakhstan, Almaty, Kazakhstan}

\end{minipage}\\
\makebox[3em]{$^{26}$}
\begin{minipage}[t]{14cm}
{\it Institute for Nuclear Research, National Academy of Sciences, and
Kiev National University, Kiev, Ukraine}

\end{minipage}\\
\makebox[3em]{$^{27}$}
\begin{minipage}[t]{14cm}
{\it Kyungpook National University, Center for High Energy Physics, Daegu,
South Korea}~$^{J}$

\end{minipage}\\
\makebox[3em]{$^{28}$}
\begin{minipage}[t]{14cm}
{\it Institut de Physique Nucl\'{e}aire, Universit\'{e} Catholique de Louvain, Louvain-la-Neuve,\\
Belgium}~$^{K}$

\end{minipage}\\
\makebox[3em]{$^{29}$}
\begin{minipage}[t]{14cm}
{\it Departamento de F\'{\i}sica Te\'orica, Universidad Aut\'onoma
de Madrid, Madrid, Spain}~$^{L}$

\end{minipage}\\
\makebox[3em]{$^{30}$}
\begin{minipage}[t]{14cm}
{\it Department of Physics, McGill University,
Montr\'eal, Qu\'ebec, Canada H3A 2T8}~$^{M}$

\end{minipage}\\
\makebox[3em]{$^{31}$}
\begin{minipage}[t]{14cm}
{\it Meiji Gakuin University, Faculty of General Education,
Yokohama, Japan}~$^{I}$

\end{minipage}\\
\makebox[3em]{$^{32}$}
\begin{minipage}[t]{14cm}
{\it Moscow Engineering Physics Institute, Moscow, Russia}~$^{N}$

\end{minipage}\\
\makebox[3em]{$^{33}$}
\begin{minipage}[t]{14cm}
{\it Moscow State University, Institute of Nuclear Physics,
Moscow, Russia}~$^{O}$

\end{minipage}\\
\makebox[3em]{$^{34}$}
\begin{minipage}[t]{14cm}
{\it Max-Planck-Institut f\"ur Physik, M\"unchen, Germany}

\end{minipage}\\
\makebox[3em]{$^{35}$}
\begin{minipage}[t]{14cm}
{\it NIKHEF and University of Amsterdam, Amsterdam, Netherlands}~$^{P}$

\end{minipage}\\
\makebox[3em]{$^{36}$}
\begin{minipage}[t]{14cm}
{\it Physics Department, Ohio State University,
Columbus, Ohio 43210, USA}~$^{A}$

\end{minipage}\\
\makebox[3em]{$^{37}$}
\begin{minipage}[t]{14cm}
{\it Department of Physics, University of Oxford,
Oxford, United Kingdom}~$^{D}$

\end{minipage}\\
\makebox[3em]{$^{38}$}
\begin{minipage}[t]{14cm}
{\it INFN Padova, Padova, Italy}~$^{B}$

\end{minipage}\\
\makebox[3em]{$^{39}$}
\begin{minipage}[t]{14cm}
{\it Dipartimento di Fisica dell' Universit\`a and INFN,
Padova, Italy}~$^{B}$

\end{minipage}\\
\makebox[3em]{$^{40}$}
\begin{minipage}[t]{14cm}
{\it Department of Physics, Pennsylvania State University, University Park,\\
Pennsylvania 16802, USA}~$^{F}$

\end{minipage}\\
\makebox[3em]{$^{41}$}
\begin{minipage}[t]{14cm}
{\it Polytechnic University, Sagamihara, Japan}~$^{I}$

\end{minipage}\\
\makebox[3em]{$^{42}$}
\begin{minipage}[t]{14cm}
{\it Dipartimento di Fisica, Universit\`a 'La Sapienza' and INFN,
Rome, Italy}~$^{B}$

\end{minipage}\\
\makebox[3em]{$^{43}$}
\begin{minipage}[t]{14cm}
{\it Rutherford Appleton Laboratory, Chilton, Didcot, Oxon,
United Kingdom}~$^{D}$

\end{minipage}\\
\makebox[3em]{$^{44}$}
\begin{minipage}[t]{14cm}
{\it Raymond and Beverly Sackler Faculty of Exact Sciences, School of Physics, \\
Tel Aviv University, Tel Aviv, Israel}~$^{Q}$

\end{minipage}\\
\makebox[3em]{$^{45}$}
\begin{minipage}[t]{14cm}
{\it Department of Physics, Tokyo Institute of Technology,
Tokyo, Japan}~$^{I}$

\end{minipage}\\
\makebox[3em]{$^{46}$}
\begin{minipage}[t]{14cm}
{\it Department of Physics, University of Tokyo,
Tokyo, Japan}~$^{I}$

\end{minipage}\\
\makebox[3em]{$^{47}$}
\begin{minipage}[t]{14cm}
{\it Tokyo Metropolitan University, Department of Physics,
Tokyo, Japan}~$^{I}$

\end{minipage}\\
\makebox[3em]{$^{48}$}
\begin{minipage}[t]{14cm}
{\it Universit\`a di Torino and INFN, Torino, Italy}~$^{B}$

\end{minipage}\\
\makebox[3em]{$^{49}$}
\begin{minipage}[t]{14cm}
{\it Universit\`a del Piemonte Orientale, Novara, and INFN, Torino,
Italy}~$^{B}$

\end{minipage}\\
\makebox[3em]{$^{50}$}
\begin{minipage}[t]{14cm}
{\it Department of Physics, University of Toronto, Toronto, Ontario,
Canada M5S 1A7}~$^{M}$

\end{minipage}\\
\makebox[3em]{$^{51}$}
\begin{minipage}[t]{14cm}
{\it Physics and Astronomy Department, University College London,
London, United Kingdom}~$^{D}$

\end{minipage}\\
\makebox[3em]{$^{52}$}
\begin{minipage}[t]{14cm}
{\it Warsaw University, Institute of Experimental Physics,
Warsaw, Poland}

\end{minipage}\\
\makebox[3em]{$^{53}$}
\begin{minipage}[t]{14cm}
{\it Institute for Nuclear Studies, Warsaw, Poland}

\end{minipage}\\
\makebox[3em]{$^{54}$}
\begin{minipage}[t]{14cm}
{\it Department of Particle Physics, Weizmann Institute, Rehovot,
Israel}~$^{R}$

\end{minipage}\\
\makebox[3em]{$^{55}$}
\begin{minipage}[t]{14cm}
{\it Department of Physics, University of Wisconsin, Madison,
Wisconsin 53706, USA}~$^{A}$

\end{minipage}\\
\makebox[3em]{$^{56}$}
\begin{minipage}[t]{14cm}
{\it Department of Physics, York University, Ontario, Canada M3J
1P3}~$^{M}$

\end{minipage}\\
\vspace{30em} \pagebreak[4]


\makebox[3ex]{$^{ A}$}
\begin{minipage}[t]{14cm}
 supported by the US Department of Energy\
\end{minipage}\\
\makebox[3ex]{$^{ B}$}
\begin{minipage}[t]{14cm}
 supported by the Italian National Institute for Nuclear Physics (INFN) \
\end{minipage}\\
\makebox[3ex]{$^{ C}$}
\begin{minipage}[t]{14cm}
 supported by the German Federal Ministry for Education and Research (BMBF), under
 contract Nos. 05 HZ6PDA, 05 HZ6GUA, 05 HZ6VFA and 05 HZ4KHA\
\end{minipage}\\
\makebox[3ex]{$^{ D}$}
\begin{minipage}[t]{14cm}
 supported by the Science and Technology Facilities Council, UK\
\end{minipage}\\
\makebox[3ex]{$^{ E}$}
\begin{minipage}[t]{14cm}
 supported by an FRGS grant from the Malaysian government\
\end{minipage}\\
\makebox[3ex]{$^{ F}$}
\begin{minipage}[t]{14cm}
 supported by the US National Science Foundation. Any opinion,
 findings and conclusions or recommendations expressed in this material
 are those of the authors and do not necessarily reflect the views of the
 National Science Foundation.\
\end{minipage}\\
\makebox[3ex]{$^{ G}$}
\begin{minipage}[t]{14cm}
 supported by the Polish Ministry of Science and Higher Education as a scientific project No.
 DPN/N188/DESY/2009\
\end{minipage}\\
\makebox[3ex]{$^{ H}$}
\begin{minipage}[t]{14cm}
 supported by the Polish Ministry of Science and Higher Education
 as a scientific project (2009-2010)\
\end{minipage}\\
\makebox[3ex]{$^{ I}$}
\begin{minipage}[t]{14cm}
 supported by the Japanese Ministry of Education, Culture, Sports, Science and Technology
 (MEXT) and its grants for Scientific Research\
\end{minipage}\\
\makebox[3ex]{$^{ J}$}
\begin{minipage}[t]{14cm}
 supported by the Korean Ministry of Education and Korea Science and Engineering
 Foundation\
\end{minipage}\\
\makebox[3ex]{$^{ K}$}
\begin{minipage}[t]{14cm}
 supported by FNRS and its associated funds (IISN and FRIA) and by an Inter-University
 Attraction Poles Programme subsidised by the Belgian Federal Science Policy Office\
\end{minipage}\\
\makebox[3ex]{$^{ L}$}
\begin{minipage}[t]{14cm}
 supported by the Spanish Ministry of Education and Science through funds provided by
 CICYT\
\end{minipage}\\
\makebox[3ex]{$^{ M}$}
\begin{minipage}[t]{14cm}
 supported by the Natural Sciences and Engineering Research Council of Canada (NSERC) \
\end{minipage}\\
\makebox[3ex]{$^{ N}$}
\begin{minipage}[t]{14cm}
 partially supported by the German Federal Ministry for Education and Research (BMBF)\
\end{minipage}\\
\makebox[3ex]{$^{ O}$}
\begin{minipage}[t]{14cm}
 supported by RF Presidential grant N 1456.2008.2 for the leading
 scientific schools and by the Russian Ministry of Education and Science through its
 grant for Scientific Research on High Energy Physics\
\end{minipage}\\
\makebox[3ex]{$^{ P}$}
\begin{minipage}[t]{14cm}
 supported by the Netherlands Foundation for Research on Matter (FOM)\
\end{minipage}\\
\makebox[3ex]{$^{ Q}$}
\begin{minipage}[t]{14cm}
 supported by the Israel Science Foundation\
\end{minipage}\\
\makebox[3ex]{$^{ R}$}
\begin{minipage}[t]{14cm}
 supported in part by the MINERVA Gesellschaft f\"ur Forschung GmbH, the Israel Science
 Foundation (grant No. 293/02-11.2) and the US-Israel Binational Science Foundation \
\end{minipage}\\
\vspace{30em} \pagebreak[4]


\makebox[3ex]{$^{ a}$}
\begin{minipage}[t]{14cm}
also affiliated with University College London,
 United Kingdom\
\end{minipage}\\
\makebox[3ex]{$^{ b}$}
\begin{minipage}[t]{14cm}
now at University of Salerno, Italy\
\end{minipage}\\
\makebox[3ex]{$^{ c}$}
\begin{minipage}[t]{14cm}
now at Queen Mary University of London, United Kingdom\
\end{minipage}\\
\makebox[3ex]{$^{ d}$}
\begin{minipage}[t]{14cm}
also working at Max Planck Institute, Munich, Germany\
\end{minipage}\\
\makebox[3ex]{$^{ e}$}
\begin{minipage}[t]{14cm}
also Senior Alexander von Humboldt Research Fellow at Hamburg University,
 Institute of Experimental Physics, Hamburg, Germany\
\end{minipage}\\
\makebox[3ex]{$^{ f}$}
\begin{minipage}[t]{14cm}
also at Cracow University of Technology, Faculty of Physics,
 Mathemathics and Applied Computer Science, Poland\
\end{minipage}\\
\makebox[3ex]{$^{ g}$}
\begin{minipage}[t]{14cm}
supported by the research grant No. 1 P03B 04529 (2005-2008)\
\end{minipage}\\
\makebox[3ex]{$^{ h}$}
\begin{minipage}[t]{14cm}
now at Rockefeller University, New York, NY
 10065, USA\
\end{minipage}\\
\makebox[3ex]{$^{ i}$}
\begin{minipage}[t]{14cm}
now at DESY group FS-CFEL-1\
\end{minipage}\\
\makebox[3ex]{$^{ j}$}
\begin{minipage}[t]{14cm}
now at DESY group FEB, Hamburg, Germany\
\end{minipage}\\
\makebox[3ex]{$^{ k}$}
\begin{minipage}[t]{14cm}
also at Moscow State University, Russia\
\end{minipage}\\
\makebox[3ex]{$^{ l}$}
\begin{minipage}[t]{14cm}
now at University of Liverpool, United Kingdom\
\end{minipage}\\
\makebox[3ex]{$^{ m}$}
\begin{minipage}[t]{14cm}
on leave of absence at CERN, Geneva, Switzerland\
\end{minipage}\\
\makebox[3ex]{$^{ n}$}
\begin{minipage}[t]{14cm}
now at CERN, Geneva, Switzerland\
\end{minipage}\\
\makebox[3ex]{$^{ o}$}
\begin{minipage}[t]{14cm}
now at Goldman Sachs, London, UK\
\end{minipage}\\
\makebox[3ex]{$^{ p}$}
\begin{minipage}[t]{14cm}
also at Institute of Theoretical and Experimental Physics, Moscow, Russia\
\end{minipage}\\
\makebox[3ex]{$^{ q}$}
\begin{minipage}[t]{14cm}
also at INP, Cracow, Poland\
\end{minipage}\\
\makebox[3ex]{$^{ r}$}
\begin{minipage}[t]{14cm}
also at FPACS, AGH-UST, Cracow, Poland\
\end{minipage}\\
\makebox[3ex]{$^{ s}$}
\begin{minipage}[t]{14cm}
partially supported by Warsaw University, Poland\
\end{minipage}\\
\makebox[3ex]{$^{ t}$}
\begin{minipage}[t]{14cm}
partially supported by Moscow State University, Russia\
\end{minipage}\\
\makebox[3ex]{$^{ u}$}
\begin{minipage}[t]{14cm}
also affiliated with DESY, Germany\
\end{minipage}\\
\makebox[3ex]{$^{ v}$}
\begin{minipage}[t]{14cm}
now at Japan Synchrotron Radiation Research Institute (JASRI), Hyogo, Japan\
\end{minipage}\\
\makebox[3ex]{$^{ w}$}
\begin{minipage}[t]{14cm}
also at University of Tokyo, Japan\
\end{minipage}\\
\makebox[3ex]{$^{ x}$}
\begin{minipage}[t]{14cm}
now at Kobe University, Japan\
\end{minipage}\\
\makebox[3ex]{$^{ y}$}
\begin{minipage}[t]{14cm}
supported by DESY, Germany\
\end{minipage}\\
\makebox[3ex]{$^{ z}$}
\begin{minipage}[t]{14cm}
partially supported by Russian Foundation for Basic Research grant
 \mbox{No. 05-02-39028-NSFC-a}\
\end{minipage}\\
\makebox[3ex]{$^{\dagger}$}
\begin{minipage}[t]{14cm}
 deceased \
\end{minipage}\\
\makebox[3ex]{$^{aa}$}
\begin{minipage}[t]{14cm}
STFC Advanced Fellow\
\end{minipage}\\
\makebox[3ex]{$^{ab}$}
\begin{minipage}[t]{14cm}
nee Korcsak-Gorzo\
\end{minipage}\\
\makebox[3ex]{$^{ac}$}
\begin{minipage}[t]{14cm}
This material was based on work supported by the
 National Science Foundation, while working at the Foundation.\
\end{minipage}\\
\makebox[3ex]{$^{ad}$}
\begin{minipage}[t]{14cm}
also at Max Planck Institute, Munich, Germany, Alexander von Humboldt
 Research Award\
\end{minipage}\\
\makebox[3ex]{$^{ae}$}
\begin{minipage}[t]{14cm}
now at Nihon Institute of Medical Science, Japan\
\end{minipage}\\
\makebox[3ex]{$^{af}$}
\begin{minipage}[t]{14cm}
now at SunMelx Co. Ltd., Tokyo, Japan\
\end{minipage}\\
\makebox[3ex]{$^{ag}$}
\begin{minipage}[t]{14cm}
now at Osaka University, Osaka, Japan\
\end{minipage}\\
\makebox[3ex]{$^{ah}$}
\begin{minipage}[t]{14cm}
now at University of Bonn, Germany\
\end{minipage}\\
\makebox[3ex]{$^{ai}$}
\begin{minipage}[t]{14cm}
also at \L\'{o}d\'{z} University, Poland\
\end{minipage}\\
\makebox[3ex]{$^{aj}$}
\begin{minipage}[t]{14cm}
member of \L\'{o}d\'{z} University, Poland\
\end{minipage}\\
\makebox[3ex]{$^{ak}$}
\begin{minipage}[t]{14cm}
now at Lund University, Lund, Sweden\
\end{minipage}\\
\makebox[3ex]{$^{al}$}
\begin{minipage}[t]{14cm}
also at University of Podlasie, Siedlce, Poland\
\end{minipage}\\

}


%% file: DESY-10-047-txt.tex
\pagenumbering{arabic} 
\pagestyle{plain}
\section{Introduction}
\label{sec-int}
The production of beauty quarks in $ep$ collisions at HERA provides a
stringent test of perturbative Quantum Chromodynamics (QCD), since the
large $b$-quark mass ($m_b~\approx~5~\gev$) provides a hard scale that
should ensure reliable predictions in all regions of phase space,
including the kinematic threshold. Especially
in this region, with $b$-quark transverse momenta comparable to or less 
than the $b$-quark mass,
next-to-leading-order (NLO) QCD calculations based on the mechanism 
of dynamical generation of the (massive) $b$ quarks
\cite{np:b412:225,*np:b454:3,*frixione3,
      np:b452:109,*pl:b353:535,pr:d57:2806}
are expected to provide accurate predictions.

The cross section for beauty production has previously been measured in 
$ep$ collisions\cite{pl:b467:156,*epj:c18:625,*pr:d70:012008,*epj:c41:453,*H1phjets,*epj:c50:1434,*pr:d78:72001,*desy-08-210, jhep:02:032,pl:b599:173,Massimo09,epj:c40:349,*epj:c45:23,*F2bH1HERAII}, 
as well as in $p\pbar$ collisions at the S$p\pbar$S
\cite{beautyUA10,*beautyUA1a,*beautyUA1,*beautyUA1b}
and Tevatron 
\cite{beautyCDF1,*beautyCDF2,*beautyCDF3,*beautyCDF4,*beautyCDF4a,*beautyCDF5,
*beautyCDF5a,*beautyCDF7,*beautyCDF8,*beautyCDF9,*beautyCDF10,
*beautyD00,*beautyD01,*beautyD02,*beautyD03}
colliders, in $\gamma \gamma$ interactions at LEP 
\cite{beautyLEP1,*beautyLEP3,beautyLEP4}, and in
fixed-target $\pi N$ \cite{WA78,*E706} and $pN$
\cite{E789,*E771,*HERAB} experiments.
Most results, including recent results from the Tevatron, are in good
agreement with QCD predictions. Some of the LEP results 
\cite{beautyLEP1}, however, deviate from the predictions.

This paper reports on a ZEUS measurement of beauty production in deep inelastic scattering (DIS) extending the kinematic region of previous ZEUS measurements \cite{pl:b599:173,Massimo09}.
The class of events investigated is
\begin{equation}\nonumber
ep\to e\;b\bbar\;X\to e\;\mathrm{jet} \; \mu\;X',
\end{equation}
in which at least one jet and one muon are found in the final state.
A data set partially overlapping with that of 
the first ZEUS measurement \cite{pl:b599:173} was used.
Looser cuts on muons and jets were applied. 
For muon identification, an extended combination of detector components was 
used. This resulted in a better detection efficiency than obtained in the 
previous analysis and allowed the threshold of the 
muon transverse momentum to be lowered. 
This is important for the extraction of the beauty contribution 
to the proton structure function, \Ftwob, for which an extrapolation to 
the full phase space has to be performed.
Such an extraction was already performed by the ZEUS 
collaboration~\cite{Massimo09} using an independent data set covering the 
kinematic range $Q^2 > 20$~GeV$^2$.
In the present analysis, the kinematic range of the measurement was extended 
to $Q^2 > 2$~GeV$^2$. A comparison to the results obtained by the H1 
collaboration\cite{epj:c40:349,*epj:c45:23,*F2bH1HERAII}, using an inclusive 
impact parameter technique, is also presented in this paper.

Due to the large $b$-quark mass, muons from semi-leptonic $b$ decays usually 
have high values of \ptrel, the transverse momentum of the muon relative 
to the axis of the jet with which they are associated. For muons from charm 
decays, from $K$ and $\pi$ decays, and in events where a hadron is 
misidentified as a 
muon, the \ptrel values are typically lower. Therefore, the fraction of events 
from $b$ decays in the data sample can be extracted by fitting the \ptrel 
distribution of the data using Monte Carlo (MC) predictions for the processes 
producing beauty, charm and light quarks. 

In this analysis, the visible cross 
section, $\sigma_{b\bar{b}}$, and differential cross sections as a function 
of $Q^2$, the transverse momentum of the muon, 
$p_T^\mu$, and its pseudorapidity\footnote{\ZpsrapB}, $\eta^\mu$, as well as 
the transverse momentum of the jet, $p_T^{\mathrm jet}$, and its pseudorapity, 
$\eta^{\mathrm jet}$, 
were measured. They are compared to leading-order (LO) plus parton-shower 
(PS) MC predictions and NLO QCD calculations.
The beauty contribution to the proton structure-function $F_2$ is extracted 
as a function of $Q^2$ and the Bjorken scaling variable, $x$, and compared to 
theoretical predictions.

\section{\label{sec-exp}Experimental set-up}
The data sample used 
corresponds to an integrated luminosity
${\cal L}=114.1\,\pm\,2.3\,\rm{pb}^{-1}$, 
collected by the ZEUS detector in the years 1996--2000.
During the 1996--97 data taking,
HERA provided collisions between an electron\footnote{Electrons and positrons are not distinguished in this paper and are both referred to as electrons.}
beam of  \mbox{$E_e=27.5\gev$} and a proton beam of
$E_p=820\gev$, corresponding to a centre-of-mass energy  
\mbox{$\sqrt s=300\gev$}
(${{\cal L}_{300}}=38.0\pm 0.6~ \rm{pb}^{-1}$). In the years
1998--2000, the proton-beam energy was
\mbox{$E_p\,=\,920\,\gev$},  corresponding to $\sqrt s=318\gev$
(${{\cal L}_{318}}=76.1\pm 1.7~\rm{pb}^{-1}$).

\Zdetdesc
\ZctddescB

\ZcaldescB

The muon system consisted of barrel, rear (B/RMUON)~\cite{brmuon}
and forward (FMUON)~\cite{zeus:1993:bluebook} tracking detectors. 
The B/RMUON consisted of 
limited-streamer (LS) tube chambers placed behind the BCAL (RCAL), inside 
and outside the magnetised iron yoke surrounding the CAL. The barrel and 
rear muon chambers covered polar angles from 34$^{\circ}$ to 135$^{\circ}$ 
and from 135$^{\circ}$ to 171$^{\circ}$, respectively. 
The FMUON consisted of six planes of LS tubes and
four planes of drift chambers covering the angular region from 
5$^{\circ}$ to 32$^{\circ}$. 
The muon system exploited the magnetic field of the iron yoke and, in the 
forward direction, of two iron toroids magnetised to  1.6\,T to provide an 
independent measurement of the muon momentum.

Muons were also detected by the sampling Backing Calorimeter (BAC) \cite{BAC}. 
This detector consisted of 5200 
proportional drift chambers which were typically 5~m long and had a wire 
spacing of 1~cm. The chambers were inserted into the magnetised iron yoke 
(barrel and two endcaps) covering the CAL. The BAC was equipped with 
analogue (for energy measurement) and digital (for muon tracking) readouts. 
The digital information from the hit wires
allowed the reconstruction of muon trajectories in two dimensions ($XY$ in 
barrel, $YZ$ in endcaps) with an accuracy of a few mm.

The luminosity was measured from the rate of the bremsstrahlung process 
$ep\to e\gamma p$. The resulting small-angle photons were measured by the 
luminosity monitor \cite{desy-92-066}, a lead--scintillator calorimeter 
placed in the HERA tunnel at $Z=-107$~m.

\section{Event selection and reconstruction}
\label{selection}

\subsection{Trigger selection}

Events containing either a scattered electron, a muon, two jets, or charmed 
hadrons were selected online by means of a three-level trigger system
\cite{zeus:1993:bluebook,uproc:chep:1992:222} through a combination of four 
different trigger chains as explained elsewhere \cite{jhep:02:032}.
The average trigger efficiency for events within 
the chosen kinematic region with a jet and with a reconstructed muon from 
$b$-quark decay was $(93 \pm 2)\%$. For events with $Q^2>20\gev^2$,
the inclusive DIS triggers yielded an efficiency of almost $100\%$. For the 
lowest $Q^2$ values, $2<Q^2<4\gev^2$, the efficiency of the combined trigger 
chains was $73\%$.

\subsection{General event selection}
Offline, the event vertex was required to be reconstructed within $|Z|< 50 \cm$
around the interaction point.
A well-reconstructed scattered electron with an impact point on the surface of 
the RCAL outside a region of $\pm12\cm$ in $X$ and $\pm6\cm$ in $Y$ around the 
beampipe and 
\begin{eqnarray}\nonumber
 E_e     &>& 10\gev \; , \\ \nonumber
 Q^2_{e} &>& 2~\gev^2
\end{eqnarray}
was required, where the estimator of $Q^2$, $Q^2_{e}$, was reconstructed using 
the energy, $E_e$, and the angle of the scattered electron.

In order to reject events from photoproduction, $Q^2 < 1$ GeV$^2$, 
the following cuts were applied:
\begin{eqnarray}\nonumber
y_{\rm JB} &>& 0.05\;, \\ \nonumber
y_{e} &<& 0.7\;, \\ \nonumber
40  &<& E-p_Z < 65 \gev \; ,
\end{eqnarray}
where $y_{\rm JB}$ and $y_{e}$ are estimators for the inelasticity, $y$, of 
the event. For small values of $y$, the Jacquet-Blondel estimator 
$y_{\rm JB}=(E-p_Z)/(2E_e)$ \cite{proc:epfacility:1979:391} was used, 
where $E-p_Z=\sum_i{E^i-p_Z^i}$ and the sum runs over all energy-flow 
objects (EFOs) \cite{epj:c1:81}. EFOs combine the information from 
calorimetry and tracking, corrected for energy loss in dead material and 
for the presence of reconstructed muons.
 
The large mass of a $b\bar{b}$ pair, at least $\approx 10\gev$, usually leads 
to a significant amount of energy deposited in the central parts of the 
detector. To reduce backgrounds from light-flavour events and charm, a cut
\begin{equation}\nonumber
E_T > 8\gev
\end{equation}
was applied, with
\begin{equation}\nonumber
  \label{eq: def_caletcorr}
 \caletcorr = \calet - {\calet}_{\left|\rm 10^{\circ} \right.} - E_T^{e}\;,
\end{equation}
where $\calet$ is the transverse energy deposited in the CAL, 
${\calet}_{\left|\rm 10^{\circ}\right.}$ is the transverse energy in a cone of 
$10^{\circ}$ around the forward beam pipe and $E_T^{e}$ is the 
transverse energy of the scattered electron.
The $b$ and $\bar{b}$ quarks also fragment and decay into a large number of 
particles. 
Therefore events with a low number of observed tracks, 
$N_\mathrm{Tracks}$, were rejected by requiring
\begin{equation}\nonumber
N_\mathrm{Tracks} \ge 8 .
\end{equation}

\subsection{Jet identification and selection}
Hadronic final-state objects were reconstructed from EFOs, which were 
clustered into jets using the $k_T$ cluster algorithm {\sc{Ktclus}} 
\cite{ktclus}
in its massive mode with the $E_T$ recombination scheme. The identified 
scattered electron was removed \cite{thesis:kahle:2006} before the clustering 
procedure, while reconstructed muons were included. 
Events were selected if they contained at least one jet with 
transverse energy, $E_T^{\rm jet}$, of  
\begin{equation}\nonumber
E_T^{\rm jet} = p_T^{\rm jet} \frac{E^{\rm jet}}{p^{\rm jet}} > 5 \gev,
\end{equation}
where $E^{\rm jet}$, $p^{\rm jet}$ and $p_T^{\rm jet}$ are the jet energy, momentum and transverse 
momentum,
and within the jet pseudorapidity ($\eta^{\rm jet}$) acceptance, 
\begin{equation}\nonumber
-2.0 < \eta^{\rm jet} < 2.5.
\end{equation}

\subsection{Muon identification and selection}

Muons were selected offline if they satisfied at least one of the following 
criteria:
\begin{itemize}
\item
a muon track was found in the inner B/RMUON chambers.
A match in position and angle to a CTD track was required.
In the bottom region, where no inner chambers are present, the outer chambers
were used instead. For muons with hits in both inner and outer chambers,
momentum consistency was required;
\item
a muon track was found in the FMUON chambers. 
Within the CTD acceptance, a match in position and angle to a CTD track was 
required and the momentum was obtained from a combined fit to the 
CTD and FMUON information.
Outside the CTD acceptance, candidates well measured in FMUON only
and fitted to the primary vertex were accepted;
\item 
a muon track or localised energy deposit was found in the BAC, and
matched to a CTD track, from which the muon momentum was obtained. In the 
forward region of the detector, an energy deposit in the calorimeter 
consistent with the passing of a minimum-ionising particle was required 
in addition in order to reduce background
related to the proton beam or to the punch through of high-energy hadrons.   
\end{itemize}
Most muons were within the geometric acceptance of more than one of these 
algorithms. 
The overall efficiency was about 80\% for muons with momenta above 2--5 GeV, 
depending on the muon pseudorapidity, $\eta^\mu$.

In the barrel region, the requirement that the muons reach at least the 
inner muon chambers implies a muon transverse momentum, $p_T^\mu$, of 
about 1.5 GeV 
or more. In order to have approximately uniform pseudorapidity acceptance, 
a cut 
$$ 
p_{T}^{\mu}>1.5\gev
$$ 
was therefore applied to all muons. 
The coverage of the tracking and muon systems resulted in
an implicit upper cutoff $\eta^{\mu}\lesssim 2.5$.
The expected signal muon distribution suggested the  
explicit cut
 $$ 
\eta^\mu > -1.6\ .
$$ 

A muon was associated with a jet if it was located within a cone of 
$\Delta R = \sqrt{\Delta\phi^2+\Delta\eta^2}< 0.7$ around the jet axis, 
where $\Delta\phi$ and $\Delta\eta$ are the distances between the muon and 
the jet in azimuth angle and pseudorapidity, respectively.
At least one muon associated with a jet was required.

After all selection cuts, the final data sample contained $19698$ events. 
In each event, only the muon candidate with the highest $p_T^\mu$ was 
considered.

\section{Monte Carlo simulation}
To evaluate the detector acceptance and to provide the signal and background 
distributions, MC samples of beauty, charm, and light flavours (LF) 
were generated, corresponding to 17, three, and about one times the 
integrated luminosity of the data, respectively. 
The beauty and charm samples were generated using the \rapgap~3 MC program 
\cite{cpc:86:147} in the massive mode ($m_c=1.5$ GeV, $m_b =4.75$ GeV), 
interfaced to {\sc Heracles} 4.6.1 \cite{cpc:69:155} 
in order to incorporate first-order electroweak corrections. 
In \rapgap, LO matrix elements are combined with 
higher-order QCD 
radiation simulated in the leading-logarithmic approximation. 
The hadronisation is simulated using the Lund string model as implemented 
in {\sc Jetset} \cite{cpc:82:74}. The lepton energy spectrum from charm 
decays was reweighted to agree with CLEO data \cite{prl:97:9}.
The lepton spectrum from beauty decays was found to be in good 
agreement \cite{thesis:kahle:2006} with that determined from $e^+e^-$ 
data.
An inclusive MC sample containing all flavours was generated in the massless 
mode using \ariadne \cite{cpc:71:15}. The subset containing only LF events 
was used for the background simulation, while the full sample was used for 
systematic studies.

The generated events were passed through a full simulation of the ZEUS 
detector based on \geant 3.13 \cite{geant}. They were subjected to the 
same trigger requirements and processed by the same reconstruction programs 
as the data.

Imperfections of the simulation of the muon range in dense materials as well 
as of the efficiency of the muon detectors were corrected 
using an independent data set of isolated muons from $\jpsi$ and 
Bethe-Heitler events \cite{thesis:bloch:2005-tmp-43fb5be7}.
Tabulated as a function of $p_T^\mu$ and $\eta^\mu$, these corrections  
were applied to MC events on an event-by-event basis.

Figure~\ref{fig-contr} shows the comparison of the MC simulation to the data 
for a selection of variables of the measured muon and the associated jet. 
The MC agrees reasonably well with the measured distributions. 
This demonstrates that the MC can be reliably used to calculate the 
detector-acceptance corrections.

\section{NLO calculations} \label{sect:theory}

Next-to-leading-order QCD predictions for the visible cross sections 
were obtained in the fixed-flavour-number scheme (FFNS) using 
\hvqdis \cite{pr:d57:2806}.
The $b$-quark mass was set to $m_b\,=\,4.75\,\gev$ and the renormalisation, 
$\mu_R$, and factorisation, $\mu_F$, 
scales to 
$\mu_R = \mu_F = \frac{1}{2}\sqrt{Q^2 + p_T^2 + m_b^2}$, where $p_T$
is the average transverse momentum of the two $b$ quarks in the Breit frame.
The parton density functions (PDF) were obtained by repeating the 
ZEUS-S \cite{zeuss} 
PDF fit in the FFNS with the quark masses set to the same values as in 
the \hvqdis calculation.

A model of $b$ fragmentation into weakly decaying hadrons and of the decay of 
$b$ hadrons into muons was used
to calculate muon observables from the partonic results. 
The hadron momentum was obtained by scaling the quark momentum according to 
the fragmentation function of Peterson et al. \cite{pr:d27:105} with the 
parameter $\epsilon=0.0035$.
The semileptonic decay spectrum for beauty hadrons was taken from
{\sc Jetset} \cite{cpc:82:74}.  
Direct ($b \to \mu$) and indirect ($b \to c(\bar c) \to \mu$ and 
$b \to \tau \to \mu$) $b$-hadron decays to muons were considered together 
according to their probabilities. 
The sum of the branching ratios of direct and indirect decays of $b$ hadrons 
into muons was fixed to $0.22$, as implemented in {\sc Jetset}\footnote{The 
small deviation from the latest 
PDG values \cite{PDG08} is negligible compared to the quoted uncertainties.}. 

The NLO QCD predictions were multiplied by hadronisation corrections
to obtain jet variables comparable to the ones used in the cross section
measurement.
These corrections are defined as the ratio of the cross sections obtained by 
applying the jet finder to the four-momenta of all hadrons and that from 
applying it to the four-momenta of all partons. They were evaluated using the 
\rapgap program; they change the NLO QCD predictions by typically 
$5\%$ or less.

The uncertainty of the theoretical predictions was evaluated by independently 
varying $\mu_R$ and $\mu_F$ by a 
factor of 2 and 1/2 and $m_b$ between $4.5$ and $5.0 \gev$. Each of these 
variations resulted in uncertainties of about 5--10\% in the kinematic 
range of this measurement.

The \hvqdis NLO predictions were also used for the extrapolation of the 
measured visible cross sections to \Ftwob. For this step, uncertainties 
on the hadronisation corrections, the branching ratios and the shape 
variation due to the choice of PDF were also included.
  
Several other predictions are available for \Ftwob. 
The predictions by the CTEQ \cite{CTEQ6} and MSTW \cite{MSTW08} groups
use NLO calculations based on the general-mass variable-flavour-number scheme 
(VFNS) with different treatments of the flavour-threshold 
region~\cite{CTEQMSTW}. 
The MSTW prediction is also available in a variant partially including 
NNLO terms \cite{MSTW08}.
The NLO prediction of GJR \cite{GJR08} is based on the FFNS.
The prediction of ABKM \cite{ABKM,*AlekhinMoch} is based on a partial 
NNLO FFNS calculation which is almost complete in the threshold region
$Q^2 \approx m_b^2$. 
Each of these calculations 
were done using PDFs extracted within the respective scheme.
The scales, masses and $\alpha_s$ values used by each prediction are 
summarised in Table \ref{tab-nlopara}.

\section{Extraction of beauty signal} \label{sect:fit}
The beauty signal was extracted from the distribution of the 
transverse momentum of the muon with respect to the momentum of the 
associated jet, \ptrel, defined as
\begin{equation} \nonumber
\ptrel=\frac{|{\vec{\boldmath{p}}^{\,\mu}}\times{\vec{\boldmath{p}}^{\rm \;jet}}|}{|\vec{\boldmath{p}}^{\rm \;jet}|},
\end{equation} 
where $\vec{p}^{\,\mu}$ is the muon and $\vec{p}^{\rm \;jet}$ the jet momentum vector. 
The fraction of beauty, $f_{\bbbar}$, and background, $f_{\rm bkg}$, 
events in the sample was obtained from a two-component fit to the shape of the 
measured \ptrel distribution, $d_\mu$, with a beauty and a background 
component:
\begin{equation} \label{Eq:fmu}
d_\mu=f_{\bbbar} d^{\bbbar}_\mu + f_{\rm bkg}d^{\rm bkg}_\mu  ,
\end{equation}
where the \ptrel distribution of beauty, $d^{\bbbar}_\mu$, was taken from 
the \rapgap MC: $d^{\bbbar}_\mu=d^{\bbbar,{\rm MC}}_\mu$. The corresponding
distribution for the background, $d^{\rm bkg}_\mu$, was obtained from the 
sum of the LF, $d_\mu^{\rm LF}$, and the charm, $d_\mu^{\ccbar}$, distributions
weighted according to the charm and LF cross sections predicted by \rapgap and 
\ariadne, respectively,
\begin{equation} \label{Eq:fbkg}
d^{\rm bkg}_\mu=rd^{\ccbar}_\mu+(1-r)d^{\rm LF}_\mu,
\end{equation}
where $r$ is the predicted charm fraction.
The distribution $d^{\rm LF}_\mu$ was obtained using 
a sample of measured CTD tracks not identified as muons.
These tracks, typically from a $\pi$ or $K$ meson, were required to 
fulfill the same momentum and angular cuts as the selected muons;
they are called unidentified tracks in the following. 
The \ptrel distribution for unidentified tracks, $d_x$, is 
expected to be similar to $d^{\rm LF}_\mu$, under the assumption that the 
probability for an unidentified track
to be identified as a muon, $P_{x\to\mu}$, does not depend strongly on \ptrel. 
Monte Carlo predictions for $d^{\rm LF}_\mu$ and $d_x$ were used to correct 
$d_x$:
\begin{equation} \label{Eq:fLF}
d_\mu^{\rm LF}={d_x}\frac{d_\mu^{\rm LF,MC}}{d_x^{\rm MC}}.
\end{equation}
The ratio $d_\mu^{\rm LF,MC}/d_x^{\rm MC}$ accounts for differences between 
$d_\mu^{\rm LF}$ and $d_x$ due to a residual \ptrel dependence of 
$P_{x\to \mu}$ and for the charm and beauty contamination in the 
unidentified track sample.

The data cannot be used to extract the distribution $d_\mu^{\ccbar}$. 
Two different options were therefore considered to describe it: 
the distribution given by the \rapgap MC, i.e. 
\mbox{$d_\mu^{\ccbar}=d_\mu^{\ccbar,{\rm MC}}$}, 
or the same
distribution corrected using the unidentified track sample, as in the case of 
the LF background: 
\begin{equation} \label{Eq:fx}
d_\mu^{\ccbar} = \frac{d_x}{d_x^{\rm MC}}{d_{\mu}^{\ccbar, {\rm MC}}}.
\end{equation} 
The average of these two distributions was taken as the 
nominal $d_{\mu}^{\ccbar}$.
The small differences between 
them were treated as a systematic uncertainty.

Figure~\ref{fig-ptrel} shows the measured distribution of the muon \ptrel
together with the results of the fit according to Eq. (\ref{Eq:fmu}). 
The fitted sum of the two components reproduces the data reasonably well. 
The fraction of beauty in the total sample is 
$f_{\bbbar}=0.16\pm0.01$~(stat.). 
For the determination of differential cross sections, the fraction of beauty 
events in 
the data was extracted by a fit performed in each cross-section bin.

The average cross sections obtained from the two different running periods 
($\sqrt{s}=$ 300 and 318 GeV) are expressed in terms of a single cross 
section at $\sqrt{s}=$ 318 GeV. The correction factor of +2\% was obtained 
using the \hvqdis NLO calculation.

\section{Systematic uncertainties}
\label{sec-sys}

The systematic uncertainties on the measured cross sections were determined 
by varying the analysis procedure or by changing the selection cuts 
within the resolution of the respective variable and 
repeating the extraction of the cross sections. The numbers given below give 
the uncertainty on the total visible cross section, $\sigma_{b\bar{b}}$. 
The systematic uncertainties on the differential distributions were determined 
bin-by-bin, unless stated otherwise. 
The following systematic studies were carried out:

\begin{itemize}
\item
 muon detection: the differences between cross sections derived 
from muons identified in the BAC and those found in the muon chambers 
was used to estimate the effect of the uncertainty in the muon detection. 
The resulting value of $\pm 7\%$ was used for all bins;
\item
fit of the beauty fraction: the uncertainty related to the signal extraction was estimated by changing the charm contribution to the background, $r$, by $+20\%$ and $-20\%$ in Eq.\;(\ref{Eq:fbkg}).
This leads to a systematic uncertainty of $^{+4}_{-3}\%$;
\item
 background \ptrel shape uncertainty: the charm \ptrel shape, $d_\mu^{\ccbar}$,
 in Eq.\;(\ref{Eq:fbkg}) was varied between the prediction from \rapgap 
and that obtained applying the correction from the unidentified track sample
in Eq.\;(\ref{Eq:fx}). 
In addition, the correction functions 
$1-\frac{d_\mu^{\rm LF,MC}}{d_x^{\rm MC}}$ and $1-\frac{d_x}{d_x^{\rm MC}}$
in Eqs. (\ref{Eq:fLF}) and (\ref{Eq:fx})
were varied by $\pm 50\%$, resulting in
a $\pm 9\%$ cross-section uncertainty; 
\item
 charm semi-leptonic decay spectrum: the reweighting to the CLEO model was 
varied by $\pm 50\%$, resulting in an uncertainty of $\pm 4\%$;
\item
 energy scale: the effect of the uncertainty in the absolute CAL energy scale of $\pm2\%$ for hadrons and of $\pm1\%$ for electrons was $^{+4}_{-5}\%$;
\item
 cut on $\calet$: a change of the cut by $\pm 1 \gev$ leads to changes in the 
cross section of $^{+2}_{-1}\%$;
\item
 cut on $N_\mathrm{Tracks}$: a change of the cut to $\ge 7$ or to $\ge 9$ 
leads to an uncertainty of $^{+2}_{-1}\%$;
\item
trigger efficiency: the uncertainty on the trigger efficiency for events with 
\mbox{$Q^2<20\gev^2$} was $\pm 2\%$.
\end{itemize}
All systematic uncertainties were added in quadrature. In addition,
a $2\%$ overall normalisation uncertainty associated with the luminosity 
measurement was added in quadrature to the uncertainty of the total cross 
section. 
This uncertainty was not included for the differential cross sections.

\section{Cross section}
\label{sec-cross}
A total visible cross section of
\begin{equation} \nonumber
\sigma_{b\bar{b}} = 70.4 \pm 5.6\; ({\rm stat.}) \pm^{11.4}_{11.3} ({\rm syst.}) \pb 
\end{equation}
was measured for the reaction $ep\to e b\bbar X\to e\,\mathrm{jet}\,\mu\ X'$ in the 
kinematic region defined by: $Q^2 > 2 \gev^2$, $0.05 < y < 0.7$, and at least 
one jet
 with $E_T^\mathrm{jet} > 5 \gev$ and $-2 < \eta^\mathrm{jet} < 2.5$ including a muon 
of $p_T^\mu > 1.5 \gev$ and $\eta^\mu > -1.6$ inside a cone 
of $\Delta R <0.7$ to the jet axis.
Jets were obtained using the $k_T$ cluster algorithm 
\ktclus \cite{ktclus} at the hadron level in its massive mode with 
the $E_T$ recombination scheme. Weakly decaying B-hadrons were treated as 
stable particles and were decayed (e.g. to a muon) only after application of 
the jet algorithm.

This result is to be compared to the \hvqdis NLO prediction of
\begin{equation} \nonumber
\sigma_{b\bar{b}}^{\rm NLO} = 46.4 \pm^{5.8}_{6.1} \pb ,
\end{equation}
where the uncertainty is calculated as described in Section \ref{sect:theory}.

Figure~\ref{fig-crq2} and Table~\ref{tab-dres} show the differential 
cross section\footnote{Cross section integrated over the bin, divided by the bin width.} 
as a function of $Q^2$ compared to the \hvqdis NLO calculation 
and the \rapgap MC prediction scaled to the data. 
Differential cross sections as functions of $p_T^\mu$, $\eta^\mu$, $p_T^{\rm jet}$
and $\eta^{\rm jet}$ are given in Fig.~\ref{fig-crmujet}. 
In shape, both the MC and the NLO QCD calculation reasonably describe the data.
The difference in normalisation is correlated to and consistent with the 
difference observed for the total cross section.
The largest fraction of the 
observed difference of about 2 standard deviations can be attributed 
to the low $x$ and $Q^2$, and therefore low $p_T$, region.

\section{Extraction of \boldmath{\Ftwob}}
\label{sec-f2b}
The beauty contribution to the proton structure-function $F_2$, \Ftwob, 
can be defined in terms of the inclusive double-differential \bbbar cross 
section in $Q^2$ and $x$ as
\begin{equation} \nonumber
\frac{d^2\sigma^{\bbbar}}{dxdQ^2}=\frac{2\pi \alpha^2}{Q^4x}\Big(\big[1+(1-y)^2\big]\Ftwob(x,Q^2)-y^2F_L^{\bbbar}(x,Q^2)\Big).
\end{equation}
The contribution from $F_L$ is small for the measured $Q^2$ and $x$ ranges 
and was neglected.
The reduced cross section for events 
containing  $b$ quarks, $\tilde{\sigma}^{b\bbar}(x,Q^2) \approx$ \Ftwob,
is defined as
\begin{equation} \nonumber
\tilde{\sigma}^{b\bbar}(x,Q^2)=\frac{d^2\sigma^{b\bbar}}{dxdQ^2}\frac{xQ^4}{2\pi\alpha^2(1+(1-y)^2)}.
\end{equation}
In this paper, the \bbbar cross section is obtained by measuring the process 
$ep\to e b\bbar X\to e\,\mathrm{jet}\,\mu\ X'$. The extrapolation from the 
measured range to the full kinematic phase 
space is performed using \hvqdis to calculate 
$\tilde{\sigma}^{b\bbar}_{\rm NLO}(x,Q^2)$. The reduced cross section is then 
determined using the ratio of the measured, 
$\frac{d^2\sigma^{b\bbar \to \mu}}{dxdQ^2}$, to calculated, 
$\frac{d^2\sigma^{b\bbar\to \mu}_{\rm NLO}}{dxdQ^2}$, double-differential 
cross sections:
\begin{equation} \label{Eq:sigred}
\tilde{\sigma}^{b\bbar}(x_i,Q^2_i)=\tilde{\sigma}^{b\bbar}_{\rm NLO}(x_i,Q^2_i)\frac{d^2\sigma^{b\bbar \to \mu}}{dxdQ^2}\bigg/\frac{d^2\sigma^{b\bbar\to \mu}_{\rm NLO}}{dxdQ^2}.
\end{equation}
The measurement was performed in bins of $Q^2$ and $x$, 
see Table~\ref{tab-diffres}.
The $Q^2$ and $x$ values for which \Ftwob was extracted, see 
Table~\ref{tab-ftwob}, were 
chosen close to the centre-of-gravity of each $Q^2$ and $x$ bin.

Predictions for $\Ftwob$ were obtained in the FFNS using \hvqdis. In 
this calculation, the same parton densities, beauty mass and factorisation and 
renormalisation scales were used as for the NLO predictions for the 
differential and double-differential cross sections discussed above.
The uncertainty of the extrapolation was estimated by 
varying the settings of the calculation
(see Section~\ref{sec-cross}) for 
$\tilde{\sigma}^{b\bbar}_{\rm NLO}(x_i,Q^2_i)$ and 
$d^2\sigma^{b\bbar\to \mu}_{\rm NLO}/dxdQ^2$
and adding the resulting uncertainties in quadrature.
The extrapolation uncertainties 
are listed in Table~\ref{tab-ftwob}.

The result of the \Ftwob extraction is shown in Fig.~\ref{fig-f2bq},
together with 
values from a previous ZEUS measurement \cite{Massimo09} focusing on the 
higher $Q^2$ region, and H1 measurements 
\cite{epj:c40:349,*epj:c45:23,*F2bH1HERAII}
using a completely different measurement technique.  
The \hvqdis + ZEUS-S NLO prediction and other predictions with different 
parameters (see Section \ref{sect:theory}) are also shown. 

The data are all compatible within uncertainties; at low $x$, the new 
measurements, in agreement with the previous ZEUS measurement, have a tendency 
to lie slightly above the H1 data.
 The largest difference is about 2 standard deviations.
 The new measurement extends the kinematic coverage down to $Q^2=3 \gev^2$
  and $x=0.00013$.
 The predictions from different theoretical approaches agree fairly well with
 each other.
 The \hvqdis predictions are somewhat lower than the ZEUS data at low $Q^2$ 
 and $x$, where the influence of the beauty-quark 
 mass is highest, while at higher $Q^2$ the data are well described by all 
 predictions.

\section{Conclusions}
\label{sec-con}
The production of beauty quarks in the deep inelastic scattering process 
$ep\to e b\bbar X\to e\,\mathrm{jet}\,\mu\ X'$ has been studied with the ZEUS
detector at HERA. Differential cross sections as a function of 
$Q^2$, $p_T^\mu$, $\eta^\mu$, $p_T^\mathrm{jet}$ 
and $\eta^\mathrm{jet}$ were measured. In all distributions, the data are 
reasonably described in shape by the Monte Carlo and by the \hvqdis NLO QCD 
calculation. 
However, at low $Q^2$ and transverse momenta, where the mass effect is largest,
\hvqdis tends to underestimate the measured values.
The extracted values of \Ftwob extend the kinematic range towards lower $Q^2$ 
and $x$ with respect to previous measurements.
They are reasonably described by different QCD predictions, whose 
spread is smaller than the current experimental uncertainty.

\section*{\label{sec-ackno}Acknowledgements}

We appreciate the contributions to the construction and maintenance of the
ZEUS detector of many people who are not listed as authors. The HERA machine
group and the DESY computing staff are especially acknowledged for their
success in providing excellent operation of the collider and the
data-analysis environment. We thank the DESY directorate for their strong
support and encouragement.

\vfill\eject

%% file: DESY-10-047-ref.tex
{
\def\bibname{\Large\bf References}
\def\refname{\Large\bf References}
\pagestyle{plain}
\ifzeusbst
  \bibliographystyle{./BiBTeX/bst/l4z_default}
\fi
\ifzdrftbst
  \bibliographystyle{./BiBTeX/bst/l4z_draft}
\fi
\ifzbstepj
  \bibliographystyle{./BiBTeX/bst/l4z_epj}
\fi
\ifzbstnp
  \bibliographystyle{./BiBTeX/bst/l4z_np}
\fi
\ifzbstpl
  \bibliographystyle{./BiBTeX/bst/l4z_pl}
\fi
{\raggedright
\bibliography{./BiBTeX/user/syn.bib,%
              ./BiBTeX/bib/l4z_articles.bib,%
              ./BiBTeX/bib/l4z_books.bib,%
              ./BiBTeX/bib/l4z_conferences.bib,%
              ./BiBTeX/bib/l4z_h1.bib,%
              ./BiBTeX/bib/l4z_misc.bib,%
              ./BiBTeX/bib/l4z_old.bib,%
              ./BiBTeX/bib/l4z_preprints.bib,%
              ./BiBTeX/bib/l4z_replaced.bib,%
              ./BiBTeX/bib/l4z_temporary.bib,%
              ./BiBTeX/bib/l4z_zeus.bib}}
}
\vfill\eject

%% file: DESY-10-047-tab.tex
\begin{table}[p]
\begin{center}
\begin{tabular}{||l|r|r|c|c|l|r||}
\hline
PDF  & Order & Scheme & \quad $\mu_F^2$ \quad  & $\mu_R^2$ & $m_b${\footnotesize (GeV)} & $\alpha_s$ \\ 
\hline\hline
{\footnotesize MSTW08 NLO}   &  $\alpha_s^2$        & {\footnotesize VFNS} \ & \multicolumn{2}{c|}{$Q^2$}                             & $\ 4.75$ & 0.1202 \\
{\footnotesize MSTW08 NNLO}  &  appr. $\alpha_s^3$  & {\footnotesize VFNS} \ & \multicolumn{2}{c|}{$Q^2$}                             & $\ 4.75$ & 0.1171 \\
{\footnotesize CTEQ6.6 NLO}  &  $\alphas,\alpha_s^2$& {\footnotesize VFNS} \ & $Q^2$ & $Q^2+m_b^2$                                     & $\ 4.5$ & 0.1180 \\
{\footnotesize GJR08 NLO}    &  $\alpha_s^2$        & {\footnotesize FFNS} \ & \multicolumn{2}{c|}{$m_b^2$}                            & $\ 4.2$ & 0.1145 \\
{\footnotesize ABKM NNLO}    & appr. $\alpha_s^3$   & {\footnotesize FFNS} \ & \multicolumn{2}{c|}{$Q^2+4m_b^2$}                       & $\ 4.5$ & 0.1129 \\ 
{\footnotesize ZEUS-S+HVQDIS}&  $\alpha_s^2$        & {\footnotesize FFNS} \ & \multicolumn{2}{c|}{$\frac{1}{4}(Q^2 + p_T^2 + m_b^2)$} & $\ 4.75$ & 0.1180\\
\hline
\end{tabular}
\caption{PDF schemes and parameters of the calculations described in Section 5 and shown in Fig. \ref{fig-f2bq}.}
 \label{tab-nlopara}
\end{center}
\end{table}

\begin{table}[p]
\begin{center}
\begin{tabular}{||c|l l l|c||}
\hline
$Q^2$ bin  & $d\sigma/d Q^2$ & $\delta_{\rm stat}$ & $\delta_{\rm syst}$&$d\sigma^{\rm NLO}/d Q^2$  \\ 
 ($\gev^2$)&                & $(\pb/\gev^2)$   &                &$(\pb/\gev^2)$ \\ 
\hline\hline 
\ \ 2 -- \ \ \ \;4 &$7.4$	        &$\pm 1.6$  	&$_{-2.4}^{+2.4}$	 &$3.4\ \ \ \ \,_{-0.3}^{+0.7}$\ \ \ \ \\
\ \ 4 -- \ \ \,10  &$3.38$	&$\pm 0.51$ 	&$_{-0.57}^{+0.56}$	 &$1.56\ \ \ _{-0.26}^{+0.21}$\ \ \ \\
\ 10 -- \ \ \,25	 &$1.10$	&$\pm 0.14$ 	&$_{-0.15}^{+0.14}$	 &$0.61\ \ \ _{-0.10}^{+0.08}$\ \ \ \\
\ 25 -- \ 100	 &$0.255$	&$\pm 0.033$	&$_{-0.036}^{+0.040}$	 &$0.163\ \,_{-0.020}^{+0.018}$\ \ \\
100 -- 1000	 &$0.0060$	&$\pm 0.0020$	&$_{-0.0016}^{+0.0016}$&$0.0092_{-0.0011}^{+0.0008}$\\
\hline\hline 
$p_T^{\mu}$  & $d\sigma/d p_T^{\mu}$ & $\delta_{\rm stat}$ & $\delta_{\rm syst}$&$d\sigma^{\rm NLO}/d p_T^{\mu}$ \\ 
 ($\gev$)&                          & $(\pb/\gev)$   &                &$(\pb/\gev)$\\ 
\hline\hline 
$1.5$ -- \ \,$2.5$	&$32.7$ 	&$\pm 4.4$		&$_{-6.0}^{+6.3}$	&\;$18.4\ \,_{-3.0}^{+2.6}$\ \ \;\\
$2.5$ -- \ \,$4.0$	&$15.4$ 	&$\pm 2.2$		&$_{-2.0}^{+2.1}$	&\;$11.9\ \,_{-1.4}^{+1.5}$\ \ \;\\
$4.0$ -- \ \,$6.0$  &\;\;$5.02$ 	&$\pm 0.90$		&$_{-0.60}^{+0.64}$	&$\ 3.66_{-0.46}^{+0.35}$\\
$6.0$ -- $10.0$ &\;\;$0.91$ 	&$\pm 0.29$		&$_{-0.14}^{+0.13}$	&$\ 0.59_{-0.07}^{+0.04}$\\
\hline\hline 
$\eta^{\mu}$  & $d\sigma/d\eta^{\mu}$ & $\delta_{\rm stat}$ & $\delta_{\rm syst}$& $d\sigma^{\rm NLO}/d\eta^{\mu}$\\ 
           &                          & $(\pb)$   &                &$(\pb)$\\ 
\hline\hline 
$-1.6$ -- $-0.5$	&\;\;$8.7$	&$\pm 2.7$		&$_{-1.6}^{+1.3}$	&$\;\;5.4_{-0.5}^{+0.8}$\\
$-0.5$ -- $\ \ 0.2$	&$16.2$		&$\pm 4.6$		&$_{-3.3}^{+3.1}$	&$16.7_{-2.5}^{+2.3}$\\
$\ \ 0.2$ -- $\ \ 0.9$	&$27.9$		&$\pm 3.6$		&$_{-4.5}^{+4.8}$	&$19.0_{-2.8}^{+2.1}$\\
$\ \ 0.9$ -- $\ \ 2.5$	&$17.1$		&$\pm 1.9$		&$_{-1.8}^{+1.8}$	&$\;\;9.4_{-1.2}^{+1.2}$\\
\hline\hline 
$p_T^{\rm jet}$  & $d\sigma/d p_T^{\rm jet}$ & $\delta_{\rm stat}$ & $\delta_{\rm syst}$& $d\sigma^{\rm NLO}/d p_T^{\rm jet}$\\ 
 ($\gev$)&                          & $(\pb/\gev)$   &                &$(\pb/\gev)$\\ 
\hline\hline 
$\ 4$ -- $10$ &$6.96$ 	&$\pm 0.75$ 	&$_{-1.52}^{+1.66}$	&$4.54_{-0.68}^{+0.64}$\\
$10$ -- $15$	&$2.69$		&$\pm 0.39$ 	&$_{-0.23}^{+0.26}$	&$2.37_{-0.28}^{+0.29}$\\
$15$ -- $30$	&$0.64$ 	&$\pm 0.14$ 	&$_{-0.07}^{+0.09}$	&$0.43_{-0.05}^{+0.03}$\\
\hline\hline 
$\eta^{\rm jet}$  & $d\sigma/d\eta^{\rm jet}$ & $\delta_{\rm stat}$ & $\delta_{\rm syst}$& $d\sigma^{\rm NLO}/d\eta^{\rm jet}$\\ 
           &                          & $(\pb)$   &                &$(\pb)$\\ 
\hline\hline 
$-1.6$ -- $-0.5$	&$14.4$		&$\pm 3.1$		&$_{-2.2}^{+2.0}$	&$\;\;6.2_{-0.5}^{+0.9}$\\
$-0.5$ -- $\ \ 0.2$	&$14.8$		&$\pm 3.8$		&$_{-2.8}^{+2.7}$	&$16.4_{-2.9}^{+2.0}$\\
$\ \ 0.2$ -- $\ \ 0.9$	&$24.0$		&$\pm 3.8$		&$_{-4.4}^{+4.4}$	&$18.2_{-2.7}^{+1.9}$\\
$\ \ 0.9$ -- $\ \ 2.5$	&$17.1$		&$\pm 2.2$		&$_{-2.2}^{+2.3}$	&$\;\;9.4_{-1.2}^{+1.3}$\\
\hline
\end{tabular}
\caption{Measured cross sections in bins of $Q^2$, $p_T^\mu$, $\eta^\mu$, $p_T^{\rm jet}$ and $\eta^{\rm jet}$ for beauty production with a muon and a jet as defined in Section \ref{sec-cross}. The statistical and systematic uncertainties are shown separately. The cross sections have an additional global uncertainty of $2\ \%$ from the luminosity uncertainty. The NLO cross sections and their uncertainties were calculated with \hvqdis.}
\label{tab-dres}
\end{center}
\end{table}

\begin{sidewaystable}[p]
\begin{center}
\begin{tabular}{||c|c|r|l l l| c||}
\hline
$Q^2$ bin  & $\log_{10}x$ bin & centre-of-gravity     &$\frac{d^2\sigma^{b\bar b \to \mu}}{d\log_{10}{x}\;dQ^2}$ & $\delta_{\rm stat}$ & $\delta_{\rm syst}$ 		&$\frac{d^2\sigma^{b\bar b \to \mu}_{\rm NLO}}{d\log_{10}{x}\;dQ^2}$  \\ 
 ($\gev^2$)&                   & $Q^2$,\ $\log_{10}x$& &$(\pb/\gev^2)$   		                  & 		&$(\pb/\gev^2)$           \\ \hline
\hline
2--4        & $-4.60$ -- $-3.50$ &   $2.86$, \ $-3.98$ & 6.3    &$\pm 1.5    $&$\pm_{1.3}^{1.4}$      &$\ \;\;2.4\pm_{0.4}^{0.4}$ \\ 
\ 4--20     & $-4.40$ -- $-3.75$ &   $6.12$, \ $-3.91$ & 0.83   &$\pm 0.17   $&$\pm_{0.13}^{0.14}$    &$\ \ 0.26\pm_{0.04}^{0.05}$ \\ 
\ 4--20     & $-3.75$ -- $-3.45$ &   $8.58$, \ $-3.65$ & 2.37   &$\pm 0.42   $&$\pm_{0.36}^{0.37}$    &$\ \ 0.83\pm_{0.13}^{0.14}$ \\ 
\ 4--20     & $-3.45$ -- $-2.50$ &  $12.45$, \ $-3.12$ & 0.80   &$\pm 0.15   $&$\pm_{0.12}^{0.12}$    &$\ \ 0.48\pm_{0.07}^{0.06}$ \\ 
20--45      & $-3.60$ -- $-3.00$ &  $28.78$, \ $-3.19$ & 0.587  &$\pm 0.086  $&$\pm_{0.073}^{0.067}$  &$\ 0.178\pm_{0.031}^{0.020}$ \\ 
20--45      & $-3.00$ -- $-1.00$ &  $32.50$, \ $-2.68$ & 0.100  &$\pm 0.034  $&$\pm_{0.024}^{0.027}$  &$\ 0.079\pm_{0.010}^{0.011}$ \\ 
\ 45--100   & $-3.30$ -- $-2.60$ &  $64.36$, \ $-2.82$ & 0.150  &$\pm 0.033  $&$\pm_{0.020}^{0.021}$  &$\ 0.067\pm_{0.007}^{0.016}$ \\ 
\ 45--100   & $-2.60$ -- $-1.00$ &  $71.74$, \ $-2.29$ & 0.045  &$\pm 0.014  $&$\pm_{0.009}^{0.011}$  &$\ 0.035\pm_{0.004}^{0.003}$ \\ 
100--250    & $-3.00$ -- $-2.30$ & $145.69$, \ $-2.49$ & 0.0206 &$\pm 0.0089 $&$\pm_{0.0067}^{0.0051}$&$\;0.0174\pm_{0.0014}^{0.0018}$ \\ 
100--250    & $-2.30$ -- $-1.00$ & $168.03$, \ $-1.99$ & 0.0054 &$\pm 0.0056 $&$\pm_{0.0024}^{0.0032}$&$\;0.0135\pm_{0.0012}^{0.0012}$ \\ 
\ 250--3000 & $-2.50$ -- $-1.00$ & $544.53$, \ $-1.73$ & 0.00065&$\pm 0.00027$&$\pm_{0.00013}^{0.00013}$&$0.00071\pm_{0.00004}^{0.00004}$ \\ \hline
\end{tabular}
\caption{Measured cross sections for different $Q^2,x$ bins for beauty production with a muon and a jet as defined in Section \ref{sec-f2b}. For each bin, the $Q^2$ and $\log_{10}{x}$ borders are shown. The centre-of-gravity, calculated to NLO using \hvqdis, is given for illustration only. The term $\frac{d\sigma}{d \log_{10} x}$ can also be read as 
$\frac{1}{x \log{10}}\frac{d\sigma}{dx}$. The statistical and systematic uncertainties are shown separately. The cross sections have an additional global uncertainty of $2\ \%$ from the luminosity uncertainty. The NLO cross sections and their uncertainties were calculated with \hvqdis.}
 \label{tab-diffres}
\end{center}
\end{sidewaystable}

\begin{table}[p]
\begin{center}
\begin{tabular}{||r|l|l l l | l||}
\hline
$Q^2$($\gev^2$)&$x$& \Ftwob  &$\delta_{\rm stat}$& $\delta_{\rm syst}$          & $\delta_{\rm extrapol}$\\ 
\hline\hline 
3 \   & $0.00013$ &  $0.0026$ & $\pm 0.0006$ & $\pm^{0.0012}_{0.0010}$  & $\pm^{0.0006}_{0.0003}$ \\ 
5 \   & $0.00013$ &  $0.0057$ & $\pm 0.0012$ & $\pm^{0.0020}_{0.0019}$  & $\pm^{0.0005}_{0.0005}$ \\ 
12 \  & $0.0002$  &  $0.0138$ & $\pm 0.0024$ & $\pm^{0.0046}_{0.0048}$  & $\pm^{0.0022}_{0.0026}$ \\ 
12 \  & $0.0005$  &  $0.0059$ & $\pm 0.0011$ & $\pm^{0.0022}_{0.0021}$  & $\pm^{0.0013}_{0.0011}$ \\ 
25 \  & $0.0005$  &  $0.0279$ & $\pm 0.0041$ & $\pm^{0.0119}_{0.0070}$  & $\pm^{0.0099}_{0.0020}$ \\ 
40 \  & $0.002$   &  $0.0101$ & $\pm 0.0034$ & $\pm^{0.0055}_{0.0055}$  & $\pm^{0.0005}_{0.0010}$ \\ 
60 \  & $0.002$   &  $0.0268$ & $\pm 0.0058$ & $\pm^{0.0096}_{0.0092}$  & $\pm^{0.0031}_{0.0019}$ \\ 
80 \  & $0.005$   &  $0.0129$ & $\pm 0.0039$ & $\pm^{0.0063}_{0.0060}$  & $\pm^{0.0008}_{0.0003}$ \\ 
130 \ & $0.002$   &  $0.0257$ & $\pm 0.0111$ & $\pm^{0.0172}_{0.0178}$  & $\pm^{0.0029}_{0.0001}$ \\ 
130 \ & $0.005$   &  $0.0061$ & $\pm 0.0063$ & $\pm^{0.0095}_{0.0093}$  & $\pm^{0.0003}_{0.0005}$ \\ 
450 \ & $0.013$   &  $0.0155$ & $\pm 0.0066$ & $\pm^{0.0099}_{0.0098}$  & $\pm^{0.0013}_{0.0002}$ \\ 
\hline
\end{tabular}
\caption{Extracted values of \Ftwob. The statistical and systematic uncertainties are shown separately. 
The uncertainty of the extrapolation to the full muon and jet phase space of the reaction $ep\to e b\bbar X\to e\,\mathrm{jet}\,\mu\ X'$ is also shown.
The cross sections have an additional global uncertainty of $2\%$ from the luminosity uncertainty.}
 \label{tab-ftwob}
\end{center}
\end{table}

%% file: DESY-10-047-fig.tex

\begin{figure}[p]
\vfill
\begin{center}
\includegraphics[width=15cm]{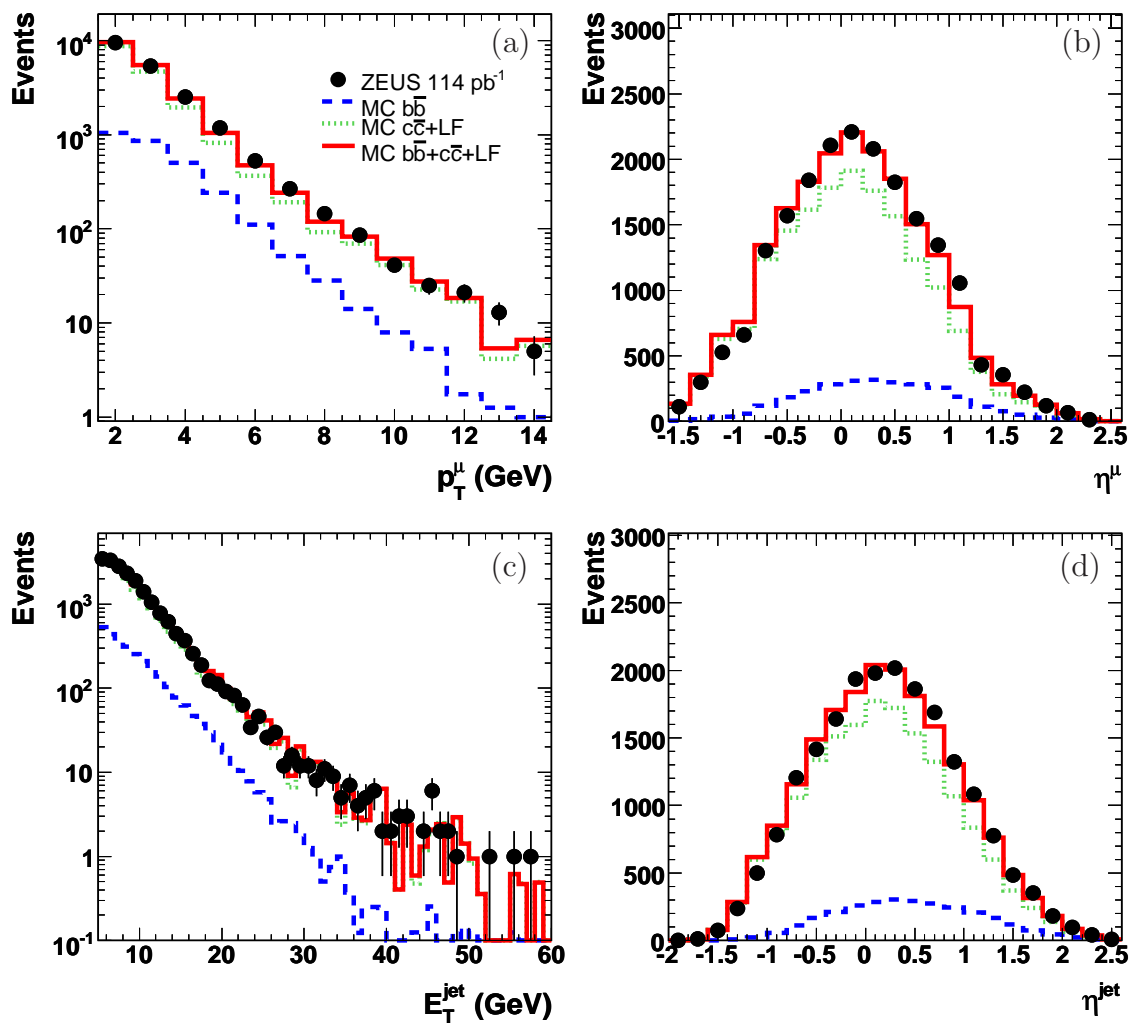}
\end{center}
\vspace{-13.7cm}
\hspace*{6.8cm} (a) \hspace*{6.7cm} (b)\vspace{5.8cm}\\

\hspace{6.8cm} (c) \hspace*{6.7cm} (d)\vspace{6.0cm}\\

\caption{Data (dots) compared to MC predictions (histograms) using the \ptrel-fit after final cuts, for which beauty (dashed), charm (dotted) and light flavours are combined (continous) as described in Section \ref{sect:fit} . The distributions of (a)~$p_T^\mu$, (b)~$\eta^\mu$, (c)~$E_T^{\rm jet}$ and (d)~$\eta^{\rm jet}$ are shown. Only statistical uncertainties are given.}
\label{fig-contr}
\vfill
\end{figure}

\begin{figure}[p]
\vfill
\begin{center}
\includegraphics[width=13.cm]{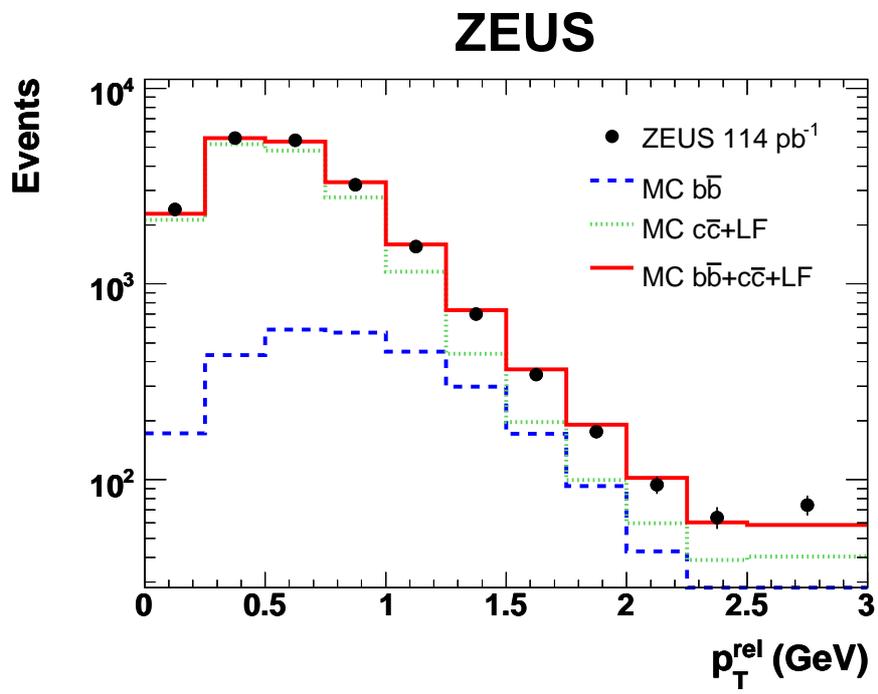}
\end{center}
\caption{Measured \ptrel-distribution and fit from MC. 
Details as in Fig. \ref{fig-contr}.}
\label{fig-ptrel}
\vfill
\end{figure}

\begin{figure}[p]
\vfill
\begin{center}
\includegraphics[width=13.cm]{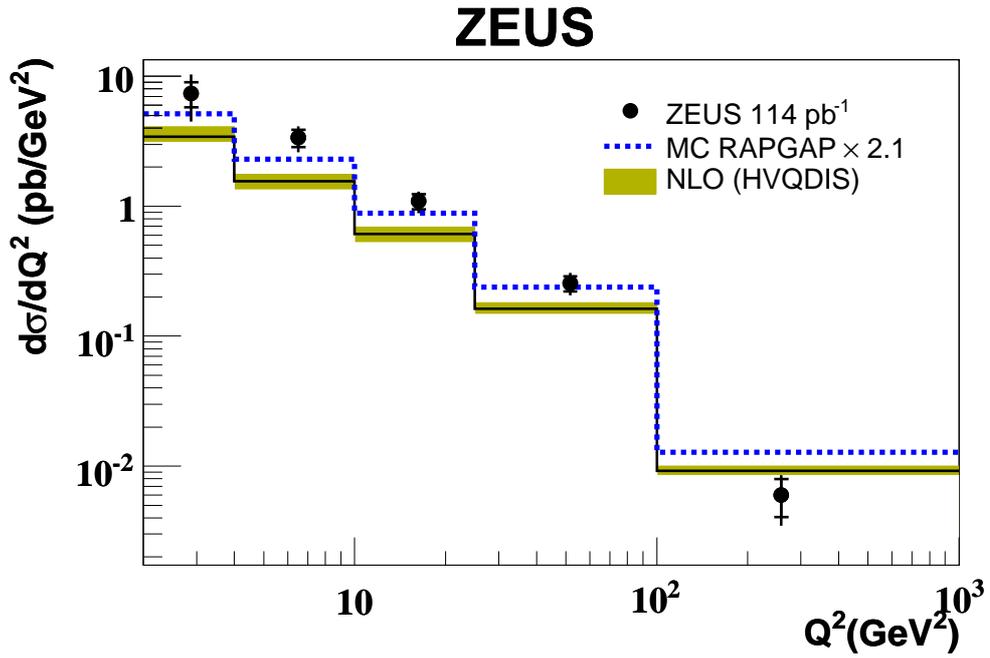}
\end{center}
\caption{Differential beauty cross section as a function of the photon virtuality, $Q^2$, for events with at least one jet and one muon, compared to the \rapgap LO+PS MC normalised to the data, and compared to the \hvqdis NLO QCD calculations. The errors on the data points correspond to the statistical uncertainty (inner error bars) and to the statistical and systematic uncertainty added in quadrature (outer error bars). The shaded bands show the uncertainty of the theoretical prediction originating from the variation of the renormalisation and factorisation scales and the $b$-quark mass.}
\label{fig-crq2}
\vfill
\end{figure}

\begin{figure}[p]
\vfill
\begin{center}
\includegraphics[width=15cm]{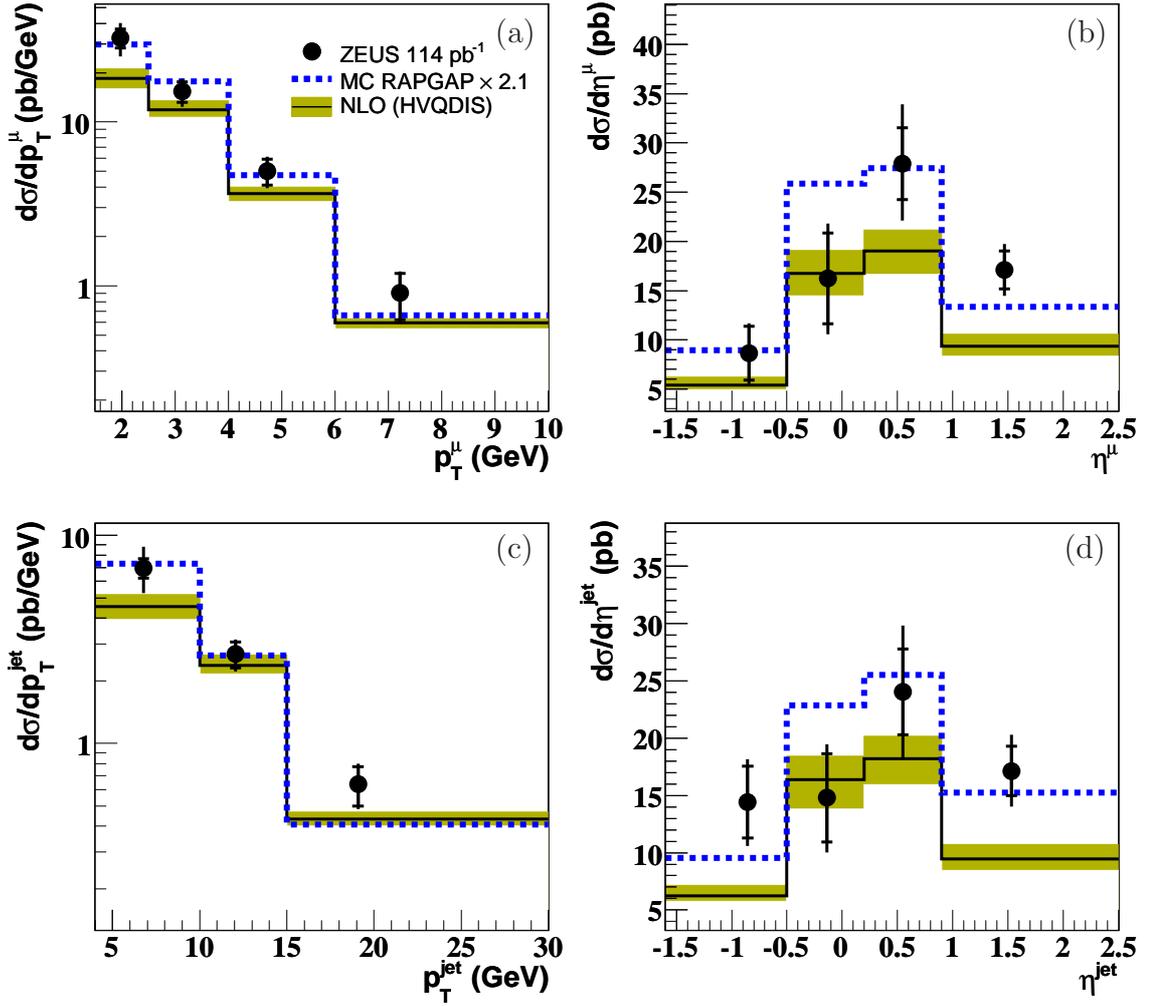}
\end{center}
\vspace{-13.7cm} 
\hspace{6.9cm} (a) \hspace*{6.7cm} (b)\vspace{5.7cm}\\

\hspace{6.9cm} (c) \hspace*{6.7cm} (d)\vspace{5.4cm}\\

\caption{Differential beauty cross section as a function of (a)~$p_T^\mu$, (b)~$\eta^\mu$, (c)~$p_T^{\rm jet}$ and (d)~$\eta^{\rm jet}$ compared to the \hvqdis NLO QCD calculations and to the scaled \rapgap MC. Other details as in Fig. \ref{fig-crq2}.}
\label{fig-crmujet}
\vfill
\end{figure}

\begin{figure}[p]
\vfill
\begin{center}
\vspace{-1cm}
\includegraphics[width=15.cm]{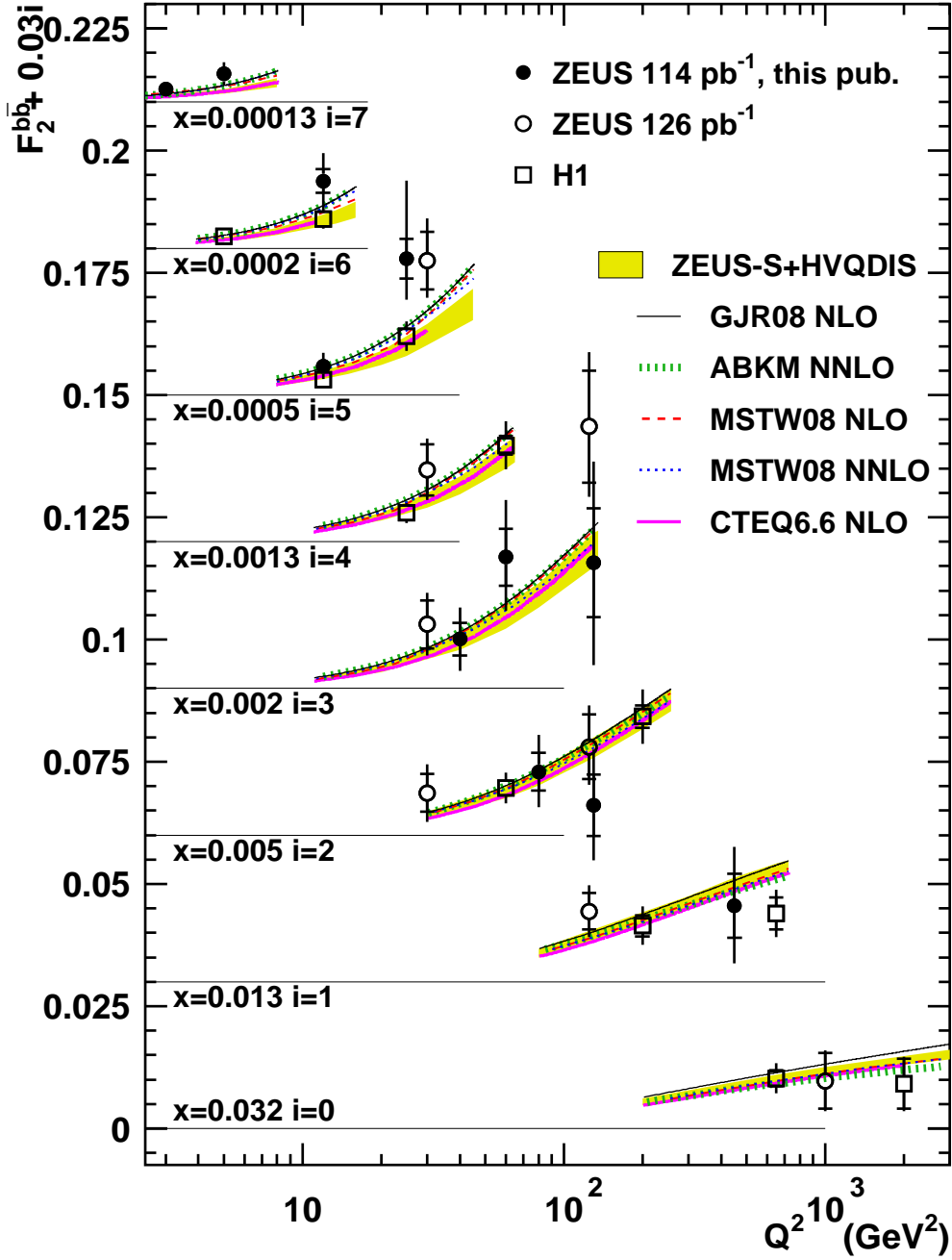}
\vspace{-0.5cm}
\end{center}
\caption{$F_2^{b\bar{b}}$ as a function of $Q^2$.  The errors on the data 
points (filled circles) correspond to the statistical uncertainty 
(inner error bars) and to the statistical and systematical uncertainty added 
in quadrature (outer error bars).
The horizontal lines indicate the zero-line for each series of measurements. 
Results from previous measurements (open symbols) and from different 
QCD predictions (lines and band) are also shown. See Section 5 and Table 1 
for details.}
\label{fig-f2bq}
\vfill
\end{figure}